\documentstyle[preprint,psfig,aps]{revtex}
\begin{document}
\draft
\title {Green Function Monte Carlo  with Stochastic Reconfiguration:\\
an effective remedy for the sign problem disease}
\author {Sandro Sorella and Luca Capriotti}
\address{Istituto Nazionale di Fisica della Materia and
International School for Advanced Studies, Via Beirut 4, 34013 Trieste, Italy}
\date{\today}
\maketitle
\begin{abstract}

A recent technique,  proposed to alleviate  the
``sign problem disease'',   is discussed in details.
As well known the ground state of a given Hamiltonian $H$
can be  obtained by applying
the imaginary time propagator $e^{-H \tau }$   to a given  trial state
$\psi_T$ for large imaginary time $\tau$ and sampling statistically the propagated
state $ \psi_{\tau} = e^{-H \tau } \psi_T$.
However the so called ``sign problem''  may appear in the simulation and
such statistical propagation
would be  practically impossible without employing  some approximation
such as the well known ``fixed node'' approximation  (FN).
This  method   allows to improve the FN dynamic
with a systematic correction scheme.
This is possible by  the simple requirement that,  after a
short imaginary time propagation via the FN  dynamic,
a number $p$  of correlation functions 
can be further constrained to be {\em exact}   by small perturbation
of the FN propagated state,  which is free of the sign problem.
By iterating this scheme  the Monte Carlo
average sign, which is almost zero when there is sign problem,
 remains stable and finite  even for large $\tau$.
The  proposed algorithm  is tested against the exact diagonalization results
available on finite lattice.
It is also shown in few test cases that
the dependence of the results   upon the few  parameters  entering the
stochastic technique can be  very easily  controlled, unless for
 exceptional cases.

\end{abstract}
\pacs{75.10.Jm,02.70.Lq,75.40.Mg}
\widetext
\section{Introduction}

In the last few years an enormous progress in the 
computational techniques has also been accompanied by better and better 
performances of modern computers. All  these developments    have  certainly 
contributed to determine  a ``feeling'' that the many body problem 
of solving a strongly 
correlated Hamiltonian,  with many electrons on a reasonably large system size,
is becoming possible with  some   computational effort. 

 The various numerical methods,  like e.g. to find 
the ground state of a physically interesting Hamiltonian, can be classified 
in two main branches  developing  from two root methods: 
the exact diagonalization  technique (ED)  and the 
variational Monte Carlo method (VMC).

The first technique is a brute force diagonalization of the   
Hamiltonian matrix, which represents a prohibitive task for large number 
of electrons as the linear dimension of this matrix grows 
exponentially with the number of electrons and the system size. 
The use of spatial symmetries and the very efficient Lanczos technique 
have  made recently possible  the exact ground state evaluation of up to 
$\sim 30$ electrons on simple lattice Hamiltonians like: the 
Heisenberg model \cite{poilblanc}, the $t-J$ model\cite{horsch},
 the Hubbard model and similar ones \cite{dagotto}.
However this system size is far from being enough for the  determination of 
the physical thermodynamic  properties of the various models.
 
A remarkable development of  the ED like methods,   is certainly the so 
called density matrix renormalization 
group technique (DMRG). In this case the ground state of a huge Hilbert  
space  Hamiltonian is sampled  by 
a small basis set  that is  iteratively improved  by using the 
renormalization group idea. 
 In one dimension this technique allows to have for instance the numerically 
exact solution of the Heisenberg spin $S=1$ model for the 
infinite size \cite{white}. 
Recently DMRG  has also been extended for high accuracy calculations on simple 
molecules \cite{martin}.

The second root of development starts  from the VMC technique \cite{vmc}. 
The VMC allows  to sample statistically a variational wavefunction
 $\psi_G(x)$, defined on a given basis set, whose elements 
 $\left\{ x \right\}$ are represented by simple {\em configurations } 
defined typically  by the electron positions and spins.  
In the most simple formulation the VMC sampling 
 can be obtained by accepting a new  configuration $x_{n+1}$ from a given 
one $x_n$ if a random number $\xi$ between zero and one satisfies  
$\xi < |\psi_G(x_{n+1})/\psi_G(x_n)|^2$, otherwise $x_{n+1}=x_n$.
This simple Metropolis algorithm generates states  $x_n$ that,
after some equilibration,   are distributed 
statistically according to the square of the variational wavefunction.  
Then  physical expectation values of operators $O^k$ -- such as pair 
correlations, electron number, total spin square, energy etc. -- 

\begin{equation} \label{vmc}
\langle O^k\rangle =
{ \langle \psi_G | O^k  | \psi_G\rangle  \over \langle  \psi_G|\psi_G\rangle } 
 = { \sum_x  \psi_G(x)^2   O^k_x \over \sum_x \psi_G(x)^2 }
\end{equation}
can be easily obtained on the given variational wavefunction,
 provided  the local estimator
$O^k_x={\langle \psi_G|O^k|x\rangle \over \langle \psi_G |x\rangle}$ of the correlation function 
$O^k$ can be computed in an efficient way.
This is typically the case since the configuration basis is particularly simple 
so that $\langle \psi_G|x\rangle$ can be easily computed, and also $O^k|x\rangle$ can be expanded 
in a few (less than the square electron number) configurations, 
for one and two body correlations. 

The iterative rule determining a  new configuration $x_{n+1}$ starting 
from a previous 
one $x_n$, and depending also on a random number, defines a Markov chain 
which  allows to obtain statistical estimates of the above 
expectation values. This is possible  even if the dimension of the Hilbert
 space is  very  large, such property 
representing  the most important  advantage of the statistical methods.  

From this point of view the Green function Monte Carlo (GFMC) 
technique \cite{ceperley} can be considered a development of 
the VMC because it allows to sample statistically  the exact 
ground state of a many body Hamiltonian $H$, instead of being restricted to 
the  variational wavefunction. 
In the GFMC the ground state is statistically sampled by a set of $M$ 
walkers $(w_i,x_i)$, $i=1,\cdots, M$, i.e., at each configuration $x_i$ is 
associated a weight  $w_i$ in order to represent a simple element $w_i\,x_i$
of the  large (or even infinite)  Hilbert space. 
In this case a Markov chain, which is  slightly more complicated 
than  the variational one,  can   be easily defined. 
As it will be shown later the new configurations and weights $(w_i,x_i)_{n+1}$ 
depend only on the previous weights and configurations  $(w_i,x_i)_n$ and $M$ 
random numbers $\xi_i$.
This iteration  is equivalent statistically to a matrix-vector product
\begin{equation} \label{iter}
\psi_{n+1}(x^\prime) = \sum_x G_{x^\prime,x} \psi_n (x) 
\end{equation}
where  $G_{x^\prime,x}$ is the lattice Green function simply related 
to the Hamiltonian matrix elements in the given basis
\begin{equation} \label{green}
G_{x^\prime,x} = \Lambda \delta_{x^\prime,x} - H_{x^\prime,x} 
\end{equation}
and  $\Lambda$ is a suitable constant, allowing the convergence of 
(\ref{iter}) to the ground state of $H$ for large $n$.
At each Markov iteration  $n$ the state $\psi_n(x)$ is sampled statistically 
 by the  walkers, which may be even a large number, but typically  
a neglectable  fraction of the total  Hilbert space dimension.

In the statistical iteration  the weights $w_i$ of the walkers 
increase or decrease exponentially so that after a few iterations most of the 
walkers have an irrelevant weight $w$ 
and some kind of reconfiguration
becomes necessary to avoid large statistical errors. 
The process to eliminate the irrelevant walkers from the statistical sampling 
is called ``branching''. This amounts for instance 
to duplicate a  walker with large $w_i$  in two walkers with half the weights 
$w_i/2$ acting on the same configuration, or drop the walkers with too small
 weights. If properly done this 
kind of process does not introduce any bias but the number of walkers is 
not constant during the corresponding Markov chain. For practical purposes  
it is necessary therefore to control the walker population number otherwise 
the simulation exceeds the maximum available memory or terminates for lack  
of walkers. This statistical reconfiguration  
instead  introduces some amount  of bias. 
 Recently a rigorous and simple way to work at finite number of walkers 
has been proposed, which simplifies the GFMC technique 
by controlling and eventually eliminating  the bias due to the finite 
number of walkers \cite{calandra}.

With a slight generalization of the previous simple technique 
it is also possible \cite{letter} to alleviate 
the ``unfamous sign problem'', which occurs 
when the matrix elements of the lattice Green function $G_{x^\prime,x}$ 
are not always positive definite.
In this case the iteration (\ref{iter}) can still have a statistical meaning 
at the price that the weights $w_i$ of the walkers are no longer restricted to 
be positive. 
It then happens that the average sign 
$\langle s\rangle_n ={ 
\langle  \sum\limits _{i=1}^M  w_i \rangle_n \over 
\langle  \sum\limits _{i=1}^M  |w_i|\rangle_n}$  
at a Markov iteration $n$  is exponentially decreasing with $n$, 
implying a dramatic decrease of the signal to noise ratio for all correlation functions. 
A remarkable improvement of the GFMC on a lattice was the extension of 
the fixed nodes (FN) approximation to lattice Hamiltonians \cite{bemmel}.
In this case the ``dangerous'' negative off-diagonal 
elements of the Green function are  neglected  
and stable simulations with always positive walker weights $w_i$ 
can be performed at the price of obtaining an approximate solution of the ground
state wavefunction.

The Green function Monte Carlo with Stochastic Reconfiguration (GFMCSR) \cite{letter}
represents a successful  attempt to improve the FN, with a stable simulation 
without any sign problem instability.   
In this scheme, better  and better approximations of the 
ground state correlation functions may be obtained, by performing controlled 
Markov chain simulations with average walker sign
 $\langle s\rangle_n$ very close to $1$ 
for each iteration $n$.  For the sake of simplicity we restrict the 
forthcoming derivation to lattice Hamiltonians but the basic ideas can be 
straightforwardly extended to the continuous case. 
 This  method is  based upon the simple
requirement that  after a
few  iterations of (\ref{iter}) via the approximate FN dynamic,
a number $p$  of correlation functions 
can be further constrained to be {\em exact} by properly 
small perturbations  of the propagated FN state $\psi^{eff}_n$, 
which is free of the sign problem.
By iterating this process the average sign  remains stable even for large $n$
and , in this limit, the method has the important property to be
 in principle {\em exact} if all possible correlation functions are included
 in this correction scheme of the FN.

In the first five sections we review the basic steps of the GFMC 
for the general case when the sign problem affects
the practical implementation of the algorithm. 
In Sec.~\ref{SR} we introduce the Stochastic Reconfiguration (SR) idea and in  
Sec.~\ref{proof}  we prove the fundamental theorem, which  justify the 
approximations used to get rid of the sign problem.
In the remaining sections we present the details of the 
algorithm and some test results, useful to  understand  how  
to implement the numerical algorithm,  for an efficient and controlled 
improvement of  the FN, even for large system sizes. 

\section{The GFMC technique}
From a general point of view the ground state $\psi_0$  of a lattice Hamiltonian 
$H$ can be obtained  by  iterating the well known power method  (\ref{iter}) 
so that $\psi_n \to \psi_0$ for large $n$, provided  the  initial state 
$\psi_T$  at  the first iteration of  Eq.~(\ref{iter}) 
($\psi_n=\psi_T$ for  $n=1$) is a trial state non-orthogonal to the
ground  state $\psi_0$.

A stochastic approach is possible if one can sample
statistically  the  matrix-vector iterations (\ref{iter}).
This is particularly important since for large systems only few 
power iterations can be applied exactly in the most fortunate cases.
The important property that allows a statistical approach is that 
 physical lattice 
Hamiltonians  are represented by very sparse matrices.
 Though the  total number of non-zero elements of $G_{x^\prime,x}$ 
is prohibitive, the number of non-vanishing entries in  each column  is  a
neglectable fraction -- of the order of the electron number -- 
of the total Hilbert space dimension. Thus  
all  the non-zero $G_{x^\prime,x}$ for fixed column  index $x$ 
can be computed even for large size.

It is therefore natural to define 
a basic element of the stochastic approach: the so called walker. 
A walker is determined by an index $x$ corresponding to a
given element $|x\rangle$ of the chosen basis and a weight $w$. 
With a stochastic approach  the walker ``walks'' in the  Hilbert 
space of the matrix $H$ and assumes a configuration $w\,x$ according 
to a given probability distribution $P(w,x)$.

The task of the GFMC approach is to define a Markov  
chain, yielding  a  probability
distribution $P_n (w,x)$  for the walker which determines 
the iterated wavefunction $\psi_n$:
\begin{equation} \label{psin}
\psi_n (x)= \langle x|\psi_n\rangle= \int dw\,w P_n(w,x)~.
\end{equation}

\section{ Importance sampling}

One of the most important advantages of the GFMC
technique is the possibility to reduce the variance of the energy by exploiting 
some information of the ground state wavefunction, known a priori on physical
grounds.  In order to understand how, we simply note that the power method 
is not restricted to symmetric matrices. 
Following Ceperley and Kalos \cite{kalos} one can  consider  in the iteration 
(\ref{iter}) not the original matrix $G$, but the slightly more involved 
non-symmetric one
\begin{equation} \label{transform} 
\bar G_{x^\prime,x } = \psi_G(x^\prime ) G_{x^\prime, x} /\psi_G(x)~,
\end{equation}
where $\psi_G$ is the so called {\bf guiding wavefunction}, that has to be as
simple as possible to be efficiently implemented in the calculation of the
matrix elements and, as we will see, as close as possible to the ground  
state of $H$.  
Here and in the following we assume that the guiding wavefunction is always 
non-vanishing for all $x$.
It is obvious that $\bar G$, though being a non-symmetric
matrix, has the same spectrum of $G$ as to any eigenvector $\psi_k(x) $ of $G$
with energy $\Lambda - E_k$ corresponds a right eigenvector of
$\bar G$ equal to $ \psi_G (x) \psi_k (x) $ with the same eigenvalue.

As shown later on, by sampling statistically the iteration (\ref{iter}) 
with $\bar G$ instead of $G$ the  walkers $(w, x)$ will be 
distributed  for large $n$ according  to $\psi_0 (x) \psi_G (x)$,
namely $\psi_{n}(x)\propto\psi_0 (x) \psi_G (x)$ in Eq.~(\ref{psin}). 
In order to evaluate 
the ground state energy, it is then enough to average the so called local energy,
\begin{equation}\label{elocal}
E_x = {  \langle \psi_G | H |x\rangle \over \langle \psi_G|x\rangle } =
 \sum_{x^\prime} \psi_G(x^\prime) H_{x^\prime,x} /\psi_G(x)~, 
\end{equation}
over the statistically sampled walkers,  because obviously:
\begin{eqnarray*}
 \langle  E_x \rangle_{\psi_0 \psi_G  } &=& 
 { \sum_x  E_x \psi_0 (x) \psi_G (x)  \over \sum_x \psi_0 (x) \psi_G (x) } \\ 
&=& { \langle \psi_0| H |\psi_G\rangle  \over  \langle \psi_0 |\psi_G\rangle } = E_0~. 
\end{eqnarray*}

Thus if $\psi_G$ is exactly equal to the ground state of $H$, by
definition $E_{x} = E_{0}$, independent of $x$, as 
$\langle \psi_G| H = E_0 \langle \psi_G| $ in (\ref{elocal}). 
This is the so called {\bf zero variance property } satisfied by the method.
Namely if the guiding wavefunction approaches  an exact eigenstate of $H$, the
method is free of statistical fluctuations.
Of course such a fortunate situation is not common at all, but by  
simply improving the guiding wavefunction  the statistical
fluctuations of the energy are much reduced, leading 
to more efficient simulations.
This property, rather obvious, is very important and non-trivial. 
Many methods, in fact, such as the path integral Monte Carlo, 
suffer of statistical fluctuations even if an exact information 
of the desired eigenstate is known. 
For Hamiltonians affected by  the sign problem it  is particularly 
important to work with  a method  which depends strongly on the quality 
of the initial guess of the ground state represented in the GFMC by
the guiding wavefunction. 
This helps a lot because 
 by  the simple and successful ``trial and error'' strategy 
one can systematically improve the guiding wavefunction and 
gain information  about the ground state. 

In general after the transformation (\ref{transform})  all mixed average 
correlation functions defined by linear operators $O^k$
\begin{equation} \label{mixav}
{\langle \psi_G| O^k  |\psi_0 \rangle \over \langle \psi_G |\psi_0\rangle} 
\end{equation}
are easily accessible by GFMC. The local estimator corresponding 
to Eq.~(\ref{mixav}) is,  analogously to (\ref{elocal}), given by
\begin{equation} \label{mixest}
O^k_x = \sum_{x^\prime}   \psi_G(x^\prime)  O^{k}_{x^\prime,x} /\psi_G (x)~, 
\end{equation}
exactly as in the variational approach (\ref{vmc}). 
This expression represents just the sum over all the possible matrix elements 
connected to $x$ of the transformed operators  $\bar O^k$ with matrix elements
\begin{equation} \label{mattra}
\bar O^{k }_{x^\prime,x} = \psi_G (x^\prime) O^k_{x^\prime,x} /\psi_G (x)~,
\end{equation}
namely  $O^k_x = \sum_{x^\prime}  \bar O^k_{x^\prime,x}$.
In order to implement  the ``importance sampling'' strategy 
it is sufficient  therefore 
to replace all the matrices involved $O^k_{x^\prime,x}$ 
including the Green function $G$ with the transformed ones $\bar O^k$ and  
$\bar G$ (\ref{transform}), 
and in all previous expressions the guiding wavefunction disappears. 
Thus the method can be considered a general method to find the maximum 
eigenvalue and eigenvector of a generic (non-symmetric) matrix $\bar G$.

In the following, for simplicity of notations, we put a bar  
over  the symbols corresponding to all the transformed matrices
(\ref{transform}) and (\ref{mattra}).

\section{Single walker formulation}
In  general  
the distribution $P_n(w,x)$ is sampled by a finite number $M$ of walkers.
 Let us first consider the simpler case
$M=1$. In order to define a statistical implementation of the matrix
multiplication  (\ref{iter}), the
standard approach is first to determine the Green function  matrix elements
$\bar G_{x^\prime,x}$ connected to $x$ which are  different
 from zero.
These matrix elements  can be generally written  in terms of three factors
\begin{equation} \label{hsign}
\bar G_{x^\prime,x}  = 
s_{x^\prime,x} p_{x^\prime,x} b_x 
\end{equation}
where   $b_x$ is a positive normalization factor, $s_{x^\prime,x}$ takes 
into account the signs of the Green function and $p_{x^\prime,x}$ is a 
stochastic matrix. All these terms will be defined explicitly below.

The basic step of the GFMC method on a lattice  is 
to define properly  the matrix
 $p_{x^\prime,x}$, because it represents the only  term in the 
decomposition (\ref{hsign})  that allows to select statistically  
only {\em one} configuration  among all the possible ones $x^\prime$ 
connected to the single  configuration $x$ of  the walker  
by the Green function application (\ref{iter}). 
Therefore $p_{x^\prime,x}$ has to represent a probability and is restricted 
to be i) normalized  $\sum_{x^\prime} p_{x^\prime,x} =1$ and ii) with 
all positive  matrix elements $p_{x^\prime,x} \ge 0$. This  is just 
the definition of a stochastic matrix (see Appendix \ref{app1}).
Since the matrix elements of $\bar G$ are  not restricted to be positive 
(sign problem)  $p_{x^\prime,x}$ is more clearly defined in terms of  an   
appropriate Green function $\bar G^{eff}$ 
with all positive matrix elements.  
Even if the latter restriction may appear rather strong, it is however 
possible that for large $n$ the approximate propagation of the state 
$\psi_n^{eff}$   by the Green function $\bar G^{eff}$ is not far from
the true propagation of $\psi_n$ by the exact Green function $\bar G$ 
in Eq.~(\ref{iter}).
$\bar G^{eff}_{x^\prime,x}$ needs  not 
to be normalized, as its normalization can be included in the 
definition of the positive constant
\begin{equation} \label{defbx}
b_x =\sum_{x^\prime} \bar G^{eff}_{x^\prime,x} 
\end{equation}
 so that
\begin{equation} \label{defgeff} 
\bar  G^{eff}_{x^\prime,x}= p_{x^\prime,x} b_x~.
\end{equation}
The typical 
choice  for $\bar G^{eff}$ is given by  the absolute value 
of the matrix elements of $\bar G$,
$\bar G^{eff}_{x^\prime,x}=|\bar G_{x^\prime,x}|$, but this is 
not the optimal choice as it will be discussed below. 

Since the  most stable right eigenvector  
$\psi_0^{eff}( x)$ of a   positive definite Green 
function -- like $\bar G^{eff}$ -- 
can be chosen positive  $\psi_0^{eff}( x) >  0$, it is important to 
implement importance sampling by the transformation (\ref{transform}) 
with a guiding wavefunction with  signs as similar as possible to the 
ones of the  ground state of $H$, so that the Green function  $\bar G$ 
has its most  stable right eigenvector $\psi_G(x) \psi_0(x) >0 $  
for most configurations $x$. In this 
case  there are good chances that the latter state  
is  well approximated by the  positive vector $\psi^{eff}_n (x)>0$, generated
 by $\bar G^{eff}$ for large $n$. 
In order to fulfill better the latter  requirement,  
here we follow a recent  development of the  FN  on a lattice, and 
we choose for $ \bar G^{eff}$ the FN Green function (with importance sampling):
\begin{equation} \label{geff}
\bar G^{eff}_{x^\prime,x} =\Lambda \delta_{x^\prime,x} - 
\bar H^{eff}_{x^\prime,x}~.
\end{equation}
The constant shift $\Lambda$ has to be large enough that all the
diagonal elements  of $\bar G^{eff}$ are strictly positive. This  is 
possible in general for the diagonal elements.  
The  full Green function  $\bar G^{eff}$ is defined in a way that the 
ground state of the Hamiltonian $H^{eff}$, 
{\em is a variational state of $H$ with an  energy better than 
 the guiding wavefunction one} \cite{ceperley1}. 
Contrary to the standard FN method, that 
neglects all the matrix elements of $H$ that cross the nodes
of the guiding wavefunction, namely the ones with $\bar H_{x\prime,x}>0$, 
we adopt here a slight modification of 
$\bar H^{eff}$ defined with non-zero matrix elements (but with opposite sign)  
when $\bar H$ has the positive ones. 
The generalization of the above  ``FN theorem'' to this case is 
straightforward and is reported in the Appendix \ref{app2}. 

The appropriate matrix elements of $\bar H^{eff}$ 
are obtained by  reversing the sign 
of  the positive  off-diagonal matrix elements of $\bar H$ and by multiplying
them by a constant $\gamma>0$
\begin{equation} \label{heffoff} 
\bar H^{eff}_{x^\prime,x}=\left\{ 
\begin{array}{ccl}
\bar H_{x^\prime,x} & {\rm if} & \bar H_{x^\prime,x}\le 0 \\
-\gamma \bar  H_{x^\prime,x} & {\rm if}
 & \bar  H_{x^\prime,x}>0
\end{array} \right.
\end{equation}
and  the diagonal ones are 
\begin{equation} \label{heffdiag}
H^{eff}_{x,x}=  H_{x,x}+(1+\gamma)  {\cal V}_{\rm sf} (x)~,
\end{equation}
where the diagonal {\sl sign-flip} contribution 
is given by\cite{ceperley1}:
\begin{equation} \label{signflip} 
{\cal V}_{\rm sf} (x) =   
\sum\limits_{\bar H_{x^\prime, x} > 0~ and~ x^\prime \ne x}
\bar{H}_{x^\prime,x}~. 
\end{equation}
Notice that there is no difference between the 
diagonal elements of the Hamiltonian $H^{eff}$ ($H$) and the ones of the
transformed  matrix $\bar H^{eff}$ ($\bar H$), as defined by 
Eq.~(\ref{mattra}). 

The equality (\ref{hsign}) holds if the factor $s_{x^\prime,x}$ 
is given by:
\begin{equation} \label{sxprimex}
s_{x^\prime,x}= \left\{ 
\begin{array}{ccl}
1 &  {\rm if} & \bar  G_{x^\prime,x}\ge 0 \\
-1/\gamma & {\rm if} &  \bar G^\prime_{x^\prime,x}<0 \\
{ \Lambda -H_{x,x} \over
  \Lambda - H^{eff}_{x,x} }
& {\rm if} & x^\prime=x 
\end{array} \right. ~.
\end{equation}

The appropriate  stochastic process 
relative to the Hamiltonian $H$ can be defined in the following three
steps, simply by allowing the weight $w$ of the walker to be also negative:
\begin{enumerate} 
\item Given the walker $(w,x)$, change the weight by scaling it with $b_x$:
$$ w \to b_x w~.$$
\item Generate randomly  a new configuration $x^\prime$ according to the
stochastic matrix $p_{x^\prime,x}$. 
\item Finally multiply the weight of the walker by
 $s_{x^\prime,x}$:
$$ w^\prime \to w s_{x^\prime,x}~.$$ 
\end{enumerate}
Without the latter step, one is actually sampling the Hamiltonian
$H^{eff}$, which we expect (or assume) to have a ground state close to the 
one of $H$, for suitably chosen guiding wavefunction.  
During the Markov iteration (MI) it is straightforward therefore to update 
both   the weight $w$ associated to the true Hamiltonian 
and the one $w^{eff}$ associated to the approximate one $H^{eff}$. 
From now on the walker will be therefore characterized by the triad:
$$ (w,w^{eff},x)~.$$ 

The previous MI  allows to define 
the evolution of the  probability density to have the walker with  weights 
$w$ and $w^{eff} >0$  in the configuration $x$, namely:
\begin{equation} \label{itersign}
P_{n+1}( w^\prime,w^{eff\,\prime}, x^\prime)= \sum_x 
{ p_{x^\prime,x} \over b_x^2  |s_{x^\prime,x}| } 
P_n ({ w^\prime \over b_x s_{x^\prime,x} }
,{ w^{eff\,\prime} \over b_x},x)~. 
\end{equation}

The first momentum   of the distribution $P$ over $w$ gives
information about the state $\psi_n (x)$ 
propagated with the exact Green function $\bar G$ and the state 
$\psi^{eff}_n (x)$ propagated with the FN Green function $\bar G^{eff}$, namely:
\begin{eqnarray} \label{states}
\psi_n (x) &=& \int d w^{eff} \int d w\,  w\, P_n (w,w^{eff},x)~,  \\
\psi^{eff}_n (x) &=& \int d w^{eff} \int d w \, w^{eff}\,  P_n (w,w^{eff},x)~. 
\end{eqnarray}
In fact it can be readily verified using (\ref{itersign}) that the 
above expressions for $\psi_n$ and $\psi^{eff}_n$, satisfy the 
iteration condition (\ref{iter}) with $\bar G$ and $\bar G^{eff}$ respectively. 

At this stage the algorithm is exact and the MI allows to sample 
the ground state  of $H$ (with sign problem) and $H^{eff}$ 
(with no sign problem) within statistical errors, that unfortunately 
may be very large, and increasing with the iteration number $n$, 
especially when there is sign problem. 

In order to have an idea on the origin of the sign problem let us discuss 
the following example. Suppose that 
$\bar H^{eff}_{x^\prime,x}= -|\bar H |_{x^\prime,x}$ 
for the off-diagonal elements and $H^{eff}=H$  otherwise.
The only information of the difference between the matrix $H$ with respect to
$H^{eff}$ is given by the sampling of the sign. In particular
it is easy to realize that in this case $w^{eff}=|w|$ at each Markov 
iteration $n$.  Then at a given iteration  $n$ 
we get 
$ \int d w^{eff} \int dw\,  w^{eff} \, P_n(w,w^{eff},x) = 
 \int dw^{eff} \int dw\,  |w| \, P_n(w,w^{eff},x) 
 \sim (\Lambda -E^{eff}_0)^n$, 
where $E^{eff}_0$ is the ground state energy
of $H^{eff}$  which is obviously below the ground state energy  $E_0$ of $H$. 
We obtain therefore the basic instability related to this Markov process, 
known as the sign problem, which, as well known, 
 is particularly difficult for fermion systems:
\begin{equation} \label{decsign} 
\langle s_n\rangle = { \sum_x \int d w^{eff}  \int dw \, w P_n(w,w^{eff}, x) 
\over \sum_x  \int d w^{eff} \int dw \, |w|\,  P_n(w,w^{eff}, x)  }  \sim 
({  \Lambda - E_0 \over  \Lambda - E_0^{eff}  } )^n.
\end{equation}
The latter relation shows that,  for  large $n$, 
 walkers with positive weight $w>0$ cancel 
almost exactly the contribution of the walkers with negative weight $w<0$ 
leaving an exponentially smaller  quantity which is obviously 
difficult to sample.
In this case only few power iterations  $n\sim 10$ are possible \cite{tklee}
and for large system size this is by far not sufficient even for 
a minor improvement of  the initial guess $\psi_G$.
It is important to emphasize that this instability does not even depend 
on the guiding wavefunction because the latter cannot change the spectrum 
of $H$ and $H^{eff}$ defined above.

By iterating several times the MI even for a single walker, 
the resulting configuration 
$(w, x)$ will be distributed according to the ground state of $H$ and by
sampling a large number of independent configurations we can evaluate for
instance the ground state energy
\begin{equation} \label{energy}
E_{0}={\langle  w E_{x} \rangle \over \langle  w \rangle}~,
\end{equation}
where the brackets $\langle \dots  \rangle$ indicate the usual stochastic average, 
namely averaging over  the independent configurations.

The configurations $x_n$ that  are generated in the Markov process are 
distributed  after many iterations  according to the  maximum right eigenstate
of the matrix $p_{x^\prime,x }$ (as, if 
we neglect the weights of the walkers, only the matrix $p$ is 
effective in the matrix product (\ref{iter})). 
This state is in general different from the  state $ \psi_G(x) \psi_{0} (x)$
we are interested in. So after many iterations the sampled configurations 
$x_n$ are distributed according to an approximate state and we can consider this 
state as a trial state $\psi_T$ for the initial iteration $n=1$ 
in the power method (\ref{iter}). At any MI $n$ we can compute  
the weight of the walker assuming that 
$L$ iterations before its value was simply $w=1$. 
In this way it is simple to compute the resulting  weight of the walker with
$L$ power Green function $\bar G$ applications:
\begin{equation} \label{defgl}  
G_{n}^L =\prod\limits_{j=1}^L b_{x_{n-j}}
s_{x_{n-j+1},x_{n-j} }~.   
\end{equation} 
Therefore for instance, in order to compute the energy with  a single Markov
chain  of many iterations,    the following quantity is usually sampled 
\begin{equation} \label{defemax}
 E_{0}= { \sum_n E_{x_n} G_n^L  \over 
\sum_n G_n^L }~, 
\end{equation}
with $L$ fixed \cite{hether}.

This would conclude the GFMC scheme, if averages over the weight variable
$G_n^L$ were possible in a stable and controlled manner.
However there are two important drawbacks for the single walker formulation. 
The first one arises because the weight $G_n^L$ of the walker grows 
exponentially with $L$ -- simply as a result of the $L$ independent 
products in Eq.~(\ref{defgl}) -- and can
assume very large values, implying diverging variances in the above averages.
This problem has a very well established solution by generalizing the 
GFMC to many walkers and introducing a scheme that enables to carry 
out walkers with reasonable values of the weights, by dropping the irrelevant 
walkers with small weights and splitting the ones with large weights. 
Recently a simple formulation of this scheme was defined at fixed number of 
walkers \cite{calandra} in a way that allows to control efficiently the residual bias 
related to the finite walker population, as discussed 
in the introduction.
The second drawback  is the more difficult one and is due to the unfamous 
sign problem.  The average sign 
$ \langle s_L\rangle  = {\sum_n G^L_n \over \sum_n  |G^L_n |}$ vanishes 
exponentially with $L$  as in Eq.~(\ref{decsign}).  
In the  formulation of Ref.~\cite{calandra} 
this problem looks quite similar 
to  the first simple one. As we will see later on, some kind of remedy can be
defined by a simple generalization of the  SR which is useful 
in the case with no sign problem. 

\section{ Carrying  many configurations simultaneously}

Given $M$ walkers we indicate the corresponding configurations 
and weights with a couple of vectors
$( \underline{w}, \underline{x} ) $, with each vector component
$(w_i,w^{eff}_{i},x_i)\,\,\,i=1,\cdots, M$ , corresponding 
to the $i^{\rm th}$ walker. Following \cite{calandra} it is 
then easy to generalize Eq.~(\ref{itersign}) to many walkers by  
the corresponding probability $P_n(\underline{w},\underline{x})$ 
of having the $M$ walkers with weights and configurations 
$(\underline{w},\underline{x})$ at the iteration $n$.
Similarly to the single walker formulation the propagated wavefunctions 
$\psi_n(x)$ and $\psi^{eff}_n(x)$   
with the true Green function $\bar G$ and the approximate one  $\bar G^{eff}$ 
read 
\begin{equation} \label{proppsimany}
\begin{array}{rcl}
\psi_n(x) &=& \int [ d \underline{w}]  \sum\limits_{ \underline{x} }
{\sum_j w_j \delta_{x,x_j}   \over M}
P_n(\underline{w},\underline{x}) \nonumber \\
\psi_n^{eff}(x) &=& \int [ d \underline{w}] \sum\limits_{ \underline{x} }
 {  \sum_j w^{eff}_{j} \delta_{x,x_j}   \over M} P_n(\underline{w},\underline{x})
\end{array}  ~,
\end{equation}
where the symbol $\int [ d  \underline{w} ]$  indicates the $2M$ 
multidimensional integral over the $(w_i,w^{eff}_{i})$ variables 
$i=1,\cdots, M$ ranging from $-\infty$ to $\infty$.
Equations (\ref{proppsimany}) are very important because they show that
the propagated quantum mechanical states $\psi_n$ and $\psi_n^{eff}$, 
which are sampled statistically, 
do not uniquely determine the walker probability function 
$P_n(\underline{w},\underline{x})$. In particular, it is perfectly possible 
 to define 
a statistical process, the  SR,  
which changes the 
probability distribution $P_n$  without changing the {\em exact}
information content, i.e., the mentioned propagated states  $\psi_n$ and $\psi_n^{eff}$.
In this way  a  linear transformation of $P_n$, described by 
a simple  kernel function 
$X(\underline{w}^\prime,\underline{x}^\prime; \underline{w}, \underline{x})$, 
will be explicitly given: 
\begin{equation} \label{pgeff}
P^\prime_n ( \underline{w}^\prime,\underline{x}^\prime)=
 \int [ d\underline{w} ] \sum\limits_{\underline{x}} 
X(\underline{w}^\prime,\underline{x}^\prime; \underline{w}
, \underline{x} )  P_n (\underline{w}, \underline{x}) ~.
\end{equation}

When there is no sign problem it is possible to define the function 
$X$\cite{calandra}  in a simple way  by requiring that 
the weights $w^\prime_j=w^{eff ^\prime}_{j}$ are 
all equal to $\sum_j w_j /M$ after the  SR.
In this case the algorithm is exact, and allows to 
perform stable simulations by applying the SR  
each few $k_p$ iterations.
Further, by increasing the number of walkers $M$, the exponential growth in  
the variance of the weights $w_j$ can be always reduced and systematically 
controlled.  In fact for large enough $M$  it is possible to work with 
$L$ sufficiently large ($L\propto M$) and obtain  results already converged 
in the power method iteration (\ref{iter}) and with small error bars.

\section{ Stochastic reconfiguration, stabilization  of the sign problem}
\label{SR}

In order to avoid the sign problem instability, at least in an approximate way,
we can follow the previous scheme as before by using
the following function $X$ that defines  the SR (\ref{pgeff})
\begin{equation} \label{reconfiguring}
X( \underline{w}^\prime,\underline{x}^\prime;\underline{w},\underline{x}) = 
\prod\limits_{i=1}^M \left( { \sum_j |p_{x_j}| 
\delta_{x^\prime_i,x_j} 
 \over \sum_j |p_{x_j}|   }  \right) \delta ( w^\prime_i - 
\beta^{-1} { \sum_j w_j \over M}  {\rm sgn}\, p_{x^\prime_i} )  
\,\, \delta ( w^{eff ~\prime}_{i} - |w^\prime_i|)~,
\end{equation}
where $\beta={\sum_j p_{x_j}\over\sum_j |p_{x_j}|} $ is the average sign
after the reconfiguration which is supposed to be much higher to stabilize the 
process.  
The kernel (\ref{reconfiguring}) has a particularly simple form since 
the outcome variables $x^\prime_j$ and $w^\prime_j$ are completely independent 
for different $j$ values. In particular it is possible to integrate easily each 
of the $M$ factors of the kernel in the variables $w^\prime_j$,
 $w^{eff \prime}_{j}$  and to sum over the configuration $x^\prime_j$, the result
being one, as it is required by the normalization condition for $P^\prime$ in 
(\ref{pgeff}).

 After the SR the exact information sampled is 
obtained by using Eq.~(\ref{proppsimany}) with $P^\prime$ instead of $P$. 
We define the corresponding quantum states $\psi^\prime_n(x)$ and  
$\psi^{eff \prime}_n(x)$,  the SR  being  exact 
whenever 
\begin{equation} \label{srexact}
\psi^\prime_n(x)=\psi_n(x) . 
\end{equation} 
 After the SR the new configurations $x^\prime_i$
are taken randomly among the old ones $\left\{ x_j \right\}$, according to the
probability ${|p_{x_i}|\over \sum_{j} |p_{x_j}|}$, defined below in terms of the given weights 
$\{ w_j \}$, $\{ w^{eff}_{j} \}$ and 
configurations $\{x_j\}$. After that the weights $w^\prime_i$ are 
changed  consistently to (\ref{reconfiguring})  
in  $w^\prime_i=\beta^{-1} {\sum_j w_j \over M} {\rm sgn}\, p_{x^\prime_i}$ and 
the FN weights, since are restricted to be positive, are defined by taking 
the absolute value of the previous  ones $w^{eff \prime}_{i}=|w^\prime_i|$.
The coefficient $\beta {\sum_j p_{x_j} \over \sum_j |p_{x_j}|}$ 
guarantees  the normalization of the two  quantum states  
after and before the reconfiguration, namely
 $\sum_x \psi_n^\prime(x) =\sum_x \psi_n (x) $. 
 This coefficient $\beta$ represents also the expected average walker  sign 
$\langle s\rangle^\prime= { \sum_j w^\prime_j   \over \sum_j |w^\prime_j| } $
  after the reconfiguration. It is supposed to be much higher than the 
 average sign before the reconfiguration $\langle s\rangle=
 { \sum_j w_j   \over \sum_j |w_j| }$, so that a stable simulation 
 with approximately constant  average sign $\langle s\rangle^\prime$ 
 can be obtained by iteratively applying the SR 
each few $k_p$ steps of the power method iteration (\ref{iter}).  

In the actual implementation of this algorithm (see Sec.~\ref{details} for the 
details) the  weights are reset to unit values after the SR:
 $w^\prime_i={\rm sgn}\, p_{x^\prime_i}$ and $w^{eff \prime}_{i}=1$, 
whereas only   the overall constant $\beta^{-1} { \sum_j w_j \over M}$,
common to all the different walkers, is stored in a sequential file. 
Then, as in the single walker formulation, at any given 
iteration $n$,  we can assume  that $L$ 
iterations before the trial state $\psi_T$ is given by the equilibrium 
distribution of walkers with unit weights $w_j= {\rm sgn } p_{x_j}$.
Therefore in order to obtain the weights 
 predicted  by the Eq.~(\ref{reconfiguring}) for $L$ power method 
iterations starting from $\psi_T$ 
it is enough to multiply the previous $L/k_p$ saved factors 
$ f_n = \beta^{-1} { \sum_j w_j   \over M}$.
This yields a natural extension  of the factors $G^L_n$ (\ref{defgl}) 
 to the many walker case
\begin{equation}
G^L_n = \prod\limits_{k=1}^{L/k_p} f_{n-k\times k_p}
\end{equation}
and the corresponding mixed average correlation functions are obtained 
by averaging the local estimators over all the iterations $n$  just 
before the SR  
(i.e.  $n$  is a multiple of $k_p$) 
\begin{equation} \label{glfactors}
 \langle  O^k\rangle= { \sum_n    G^L_n { \sum_j w_j  O^k_{x_j}  } \over 
\sum_n  G^L_n { \sum_j w_j } }~,
\end{equation} 
where,  in the above equation, the weights $w_j$
and the local estimators  $O^k_{x_j} $   are evaluated only  
  before the SR. 

The only left quantity to define properly the whole algorithm 
consistently with Eq.~(\ref{reconfiguring})  are the important 
coefficients  $p_{x_j} $ which {\em  have not }  to be  
assumed positive. 
These coefficients may depend on all the weights $w_j$, the configurations 
$x_j$ and the FN weights $w^{eff}_{j}$. 

The  choice $p_{x_j}=w_j$ is  exact in the sense that 
$\psi_n^\prime (x)=\psi_n(x)$,   and  
coincides with the one  for the case with 
no sign problem \cite{calandra}.
 However this choice is obviously  
not convenient, because this reconfiguration will not improve the sign, which
will decay exponentially in the same way. 

 Instead, in the case with sign problem, 
 we can  parameterize the coefficients
$p_{x_j}$ by assuming they are  close enough to the   positive definite 
weights  $\{ w^{eff}_{j} \}$, the ones obtained with the FN 
Green function $G^{eff}$. 
The rational of this choice is that, though the weights $w^{eff}_{j}$ may be 
occasionally very different from the exact weights $w_j$ -- namely 
the sign can be wrong -- they sample a state $\psi^{eff}_n(x)$ which is 
supposed to be quite close to the exact propagated state $\psi_n(x)$. This 
condition is clearly verified for an appropriate choice 
of the guiding wavefunction $\psi_G$, which makes   the FN accurate. 
Then we assume that small perturbations over the state $\psi^{eff}_n(x)$   
may lead to fulfill the equality  (\ref{srexact}) with an arbitrary small error. 
In the case with sign problem in fact, we release the exact SR 
condition (\ref{srexact}) 
to be satisfied within some error. This error will affect the equilibrium 
walker  distribution $P_n$ for large $n$, but there will be no problem if 
this error  i) is small and ii) can be  reduced within the desired 
accuracy. 

In the most simple and practical formulation we require that the average 
energy before and after the SR  coincide
\begin{equation} \label{eqe}
\sum_{x^\prime,x} \bar{H}_{x^\prime,x} \psi_n(x) = 
\sum_{x^\prime,x} \bar{H}_{x^\prime,x} \psi^\prime_n(x) 
\end{equation}
(the denominators in the mixed averages (\ref{mixav}) are already equal 
by definition as  $\sum_x \psi_n(x)= \sum_x \psi^\prime_n (x)$ 
for the chosen $\beta$ in (\ref{reconfiguring})).
Then we define
 $$p_{x_j} = w^{eff}_{j}  ( 1 + \alpha (  E_{x_j} - \bar E_{eff}) )$$
and
\begin{equation}
\begin{array}{rcl} 
\bar E_{eff}&=& { \sum_j  w^{eff}_{j}  E_{x_j}  \over \sum_j w^{eff}_{j} }
  \nonumber \\
\bar E &=& {\sum_j w_j E_{x_j} \over \sum_j w_j } 
\end{array}~,
\end{equation}
where  $E_{x_j}$ is the local energy (\ref{elocal}) 
associated to the configuration $x_j$.
Thus  $\bar E$ represents the estimate of the average energy correctly sampled 
with the sign, whereas $\bar E_{eff}$ is the one with no sign problem. 
In order to satisfy the requirement (\ref{eqe})  we just determine $\alpha$ by
\begin{equation} \label{defalpha}
\alpha= { \bar E - \bar E_{eff}  \over \bar E_{eff}^2  - (\bar E_{eff} )^2 } 
\end{equation}
where $\bar E_{eff}^{2} = { \sum_j  w^{eff}_{j}  E_{x_j}^2 \over \sum_j w^{eff}_{j} } $
is the average square energy over the positive distribution $w^{eff}_{j}$. 
 
A simple calculation shows that with this 
reconfiguration,  that clearly improves the sign, the value of the energy 
(the mixed average energy) remains statistically the same before and after 
the SR (see next Section and Appendix \ref{app3}).
It is clear however that this is not enough to guarantee convergence to the 
exact ground state, because fulfillment of (\ref{eqe}) does not imply the 
exact equality (\ref{srexact}).
We can improve  the definition of the constants $p_{x_j}$  
by including an arbitrary  number $p$ of parameters with   $p <<M$
\begin{equation}  \label{defpx}
p_{x_j} = w^{eff}_j ( 1+  \alpha_1 ( O_{x_j}^1 -\bar O_{eff}^1)
+ \cdots + \alpha_p (O_{x_j}^p -\bar O_{eff}^p) ) 
\end{equation}  
proportional to the fluctuations 
$O_{x_j}^k - \bar O_{eff}^k$  of   $p$ different 
operators $O^k$ with corresponding local estimators  
$O_{x_j}^k = { \langle \psi_G | O^k |x_j\rangle \over \langle \psi_G
|x_j\rangle } $ for $k=1,\cdots, p$, and average value over the positive weights:
$ \bar O_{eff}^k={\sum_j w^{eff}_{j} O^k_{x_j} \over \sum w^{eff}_{j} } $. 
With the more general form (\ref{defpx}) for the coefficients $p_{x_j}$ 
it is possible to fulfill that all the mixed averages for the chosen 
$p$ operators -- not only the energy --  have the same value 
before and after the SR:
\begin{equation} 
\label{srcon}
\sum_{x^\prime,x} \bar O^k_{x^\prime,x} \psi_n(x) = 
\sum_{x^\prime,x} \bar O^k_{x^\prime,x} \psi^\prime_n(x)~. 
\end{equation} 
In general the reference weights $w^{eff}_j$ in Eq.~(\ref{defpx})
may be also different from the ones generated by the FN Green function, 
the only restriction is that $w^{eff}_j >0$ 
for each walker $j$ (see Appendix \ref{app3}). 

It is proven in the next Section that in order to fulfill 
exactly the SR conditions (\ref{srcon}) it is {\em sufficient }  
that the coefficients  $p_{x_j}$  are chosen in a way that
\begin{equation} \label{condition}
{ \sum\limits_j p_{x_j} O_{x_j}^k   \over \sum\limits_j p_{x_j} } = 
{ \sum\limits_j  w_j O_{x_j}^k \over \sum\limits_j  w_j }~,
\end{equation} 
which can be fulfilled with a solution of a simple linear system for the 
unknown variables $\alpha_k$, for $k=1,\cdots, p$, as described in the Appendix  
\ref{app3}.
The conditions (\ref{condition}) are much simpler  
because they can be satisfied  at  a given  iteration of the Markov process.
The  theorem, proven in the next section, guarantees that the exact
(\ref{srcon}) are implied by the constraints (\ref{condition})  
after the complete statistical average over the probability walker
distribution $P_n$. 

Thus, asymptotically, by adding more and more
parameters $\{\alpha_j \}$, we can achieve
$\psi^\prime_{n}(x)=\psi_{n}(x)$ strictly, since the
distribution $\psi_{n}(x)$ 
is completely determined by its  correlation functions. 
The  proof of this important statement is very simple. Consider first the
diagonal operators. All these operators may be written as linear combinations of
the ``elementary'' ones 
$O^{x_0}_{x^\prime,x}=\delta_{x^\prime,x} \delta_{x,x_0}$
acting on  a single  configuration
$x_0$,  plus at most some constants.  If 
conditions (\ref{srcon}) are satisfied  for  {\em all}  the  elementary
operators it immediately follows that  $\psi^\prime_n  (x_0)= \psi_n(x_0) $ for
all $x_0$, which is the  exact SR condition (\ref{srexact}). 

Then it is simple to show that the coefficients
$p_{x_j}$, determining $P^\prime_n$ and $\psi^\prime_n$, 
are invariant  for any constant shift   
of the  operators $O^k$.  Further with  a little algebra it
turns out that  these coefficients $p_{x_j}$ do not change  for any
arbitrary   linear transformation of the chosen operator set:  $O^{k \prime}  =
\sum_{k} L_{k^\prime,k} O^k$ (with real $L$ and $\det L \ne 0$) 
(see Appendix \ref{subapp3}). 
Thus the proven convergence of the GFMCSR is obtained for any sequence of  
diagonal  operators, that, with increasing $p$, becomes complete. 
For non-diagonal operators $O_{x^\prime,x}$ note simply that they assume 
the same mixed average values  of the equivalent diagonal ones
$O^{diag}_{x^\prime,x}=\delta_{x^\prime,x}  \sum_{x^\prime} O_{x^\prime,x}$. 
Thus the proof   that GFMCSR  converges in principle to the
exact solution is valid in general even when  non-diagonal operators,
such as  the Hamiltonian itself for the energy,  are included  in the 
conditions (\ref{srcon})  $\Box$.

\section{Formal proof of the  GFMCSR conditions } 
\label{proof}

As stated before the SR conditions (\ref{srcon}) read
\begin{equation}
\label{srconn}  \sum_{x^\prime,x} \bar  O^k_{x^\prime,x} \psi^\prime_n(x)= 
\sum_{x^\prime,x} \bar  O^k_{x^\prime,x}  \psi_n (x)~,
\end{equation} 
for $k=1,\cdots,  p$, with  the 
normalization one $\sum_x \psi^\prime_n (x)=\sum_x  \psi_n(x)$.

The wavefunction $\psi^\prime_n (x)$ after the 
SR conditions defined by (\ref{reconfiguring})  
can be explicitly written in terms of the original walker probability 
distribution. To this purpose  we single out in the definition of 
$\psi^\prime_n (x)$  
\begin{equation}
\psi_n^\prime(x)=\int [d\underline{w}^\prime ] \sum_{\underline{x}^\prime } 
P^\prime_n (\underline{w}^\prime,\underline{x}^\prime) { \sum_j 
\delta_{x,x^\prime_j} w^\prime_j \over M},
\end{equation}
a term $k$ in the above summation over $j$ which gives an additive 
contribution to $\psi^\prime_n$, namely 
$\psi^\prime_n= { 1 \over M} \sum_k \left\{ \psi^\prime_n \right\}_k $ 
with
\begin{equation}
\left\{ \psi_n^\prime(x) \right\}_k =
\int [d\underline{w}^\prime ] \sum_{\underline{x}^\prime }
\int [d\underline{w} ] \sum_{\underline{x}} 
X(\underline{w}^\prime,\underline{x}^\prime;\underline{w},\underline{x})  
P_n (\underline{w},\underline{x}) 
\delta_{x,x^\prime_k} w^\prime_k~,
\end{equation}
where in  the above equation  we have substituted  
the definition of $P^\prime$ in terms of $P$ given by
Eqs.~(\ref{pgeff}) and (\ref{reconfiguring}).
In the latter equation it is easy to integrate over all variables 
$w^\prime_j, w^{eff ~ \prime}_{j},x^\prime_j$ for $j \ne k$ using that
 the kernel $X$ is particularly simple as previously discussed. 
Then,  the remaining three  integrals and summations over
$w^\prime_k, w^{eff ~ \prime}_{k},x^\prime_k$ can be easily performed using 
the simple $\delta$ functions which appear  in the kernel $X$ and 
the definition of $\beta={ \sum_j p_{x_j} \over \sum_j |p_{x_j}| }$, so that 
one easily obtains
\begin{equation}
\left\{ \psi_n^\prime(x) \right\}_k = 
\int [d\underline{w} ] \sum_{\underline{x}}
P_n (\underline{w},\underline{x}) { \sum_j w_j \over M} {\rm sgn} p_{x} 
{\sum_j |p_{x_j}| \delta_{x,x_j}  \over  \sum_j p_{x_j} }~.
\end{equation}
It is important to remark  that,  in the above equation, 
the sign of $p_x$ ($ {\rm sgn} p_{x}$) 
depends only on the configuration  $x$ chosen among the old configurations 
$x_j$, determining the vector $\underline{x}$ in 
$P_n (\underline{w},\underline{x}).$  In particular if there are more
 walkers acting on the same configuration ($x_j=x$  for more than one $j$)  
${\rm sgn} p_{x}$ is the same for all the corresponding 
indices, as implied by the definition (\ref{defpx}) of $p_{x_j}$  
and the condition $w^{eff}_j > 0$ valid for all $j$.
We can therefore replace in general 
$ {\rm sgn} p_{x} {\sum_j |p_{x_j}| \delta_{x,x_j}  \over  \sum_j p_{x_j} } = 
{ \sum_j p_{x_j} \delta_{x,x_j}  \over  \sum_j p_{x_j} } $
and obtain the closed expression for $\psi^\prime_n(x)$ after the simple 
summation on the index $k$:  
\begin{equation} 
\psi^\prime_n(x) = \int [d \underline{w} ] \sum_{\underline{x}} 
P_n(\underline{w},\underline{x}) ( { \sum_j w_j \over M} ) {\sum_j p_{x_j}
\delta_{x,x_j}  \over \sum_j p_{x_j} }~. 
\end{equation}

Then the  normalization condition $\sum_x \psi^\prime_n(x){=}
\int [d \underline{w} ] \sum_{\underline{x}} 
P_n(\underline{w},\underline{x}) ( { \sum_j w_j \over M} ) 
= \sum_x \psi_n(x)$ 
easily follows.
On the other hand  the left hand side of Eqs.~(\ref{srcon}) can be 
also computed easily, yielding 
\begin{equation} \label{blah}
\sum_{x^\prime,x}  \bar  O^k_{x^\prime,x} \psi^\prime_n(x)=
\int [d \underline{w} ] \sum_{\underline{x}} 
P_n(\underline{w},\underline{x}) ( { \sum_j w_j \over M} ) {\sum_j p_{x_j}
  O^k_{x_j}  \over \sum_j p_{x_j} }~,
\end{equation}
where $O^k_{x_j}=\sum_{x^\prime} \bar O_{x^\prime,x_j}$ is the 
mixed estimator of the operator $O^k$.

Finally, by substituting  the conditions (\ref{condition}) into 
the previous equation, one obtains
\begin{equation} 
\sum_{x^\prime,x} \bar  O^k_{x^\prime,x} \psi^\prime_n(x)= 
\int [d \underline{w} ] \sum_{\underline{x}} 
P_n(\underline{w},\underline{x})   {\sum_j w_j
O^k_{x_j}  \over M } = \sum_{x^\prime,x} \bar O^k_{x^\prime,x} \psi_n(x)~, 
\end{equation}
which proves the statement at the beginning of this section.

\subsection{ Optimization of the weights }
The definition of the weights $p_{x_j}$ that satisfies  the SR 
conditions (\ref{srcon}) is highly arbitrary 
because as we have mentioned before the probabilities $P_n$ and $P^\prime_n$ 
do not uniquely determine the quantum states $\psi_n$ and $\psi^\prime_n$ 
that are subject to the conditions (\ref{srcon}).
In this sense  there may be different definitions of the weights $p_{x_j}$ 
that may behave differently at finite $p$ with  less or more accuracy. 
Though Eqs.~(\ref{srcon}) 
are equally satisfied for different choices  of   the coefficients  $p_{x_j}$ the two states 
$\psi_n$ and $\psi^\prime_n$ may be much closer (less bias) for an optimal 
choice.  
 The optimal choice that minimizes the distance 
$ | \psi_n - \psi^\prime_n|$, at fixed number $p$ of correlation functions
 included in the SR,   has not probably been found yet. We have attempted 
several choices for the reference weights 
$w^{eff}_j$ of Eq.~(\ref{defpxa}) (with $w^{f}_j=w^{eff}_j$) 
but until now no significant improvement of the  simplest 
FN ones \cite{letter} has been obtained.

\section{Details of the algorithm} \label{details}

In this section the flow chart of the GFMCSR algorithm is briefly sketched.
As described in the Appendix \ref{app4} it is possible to  work without
the extra constant shift $\Lambda$ and
apply directly $e^{-H \tau }$, the usual imaginary time propagator, to
filter out the ground state from the chosen trial wavefunction $\psi_T$.

For practical purposes, the algorithm can be divided into
three steps, 1) the Green function (GF) evolution, 2) the SR
and 3) the measurements of physical mixed average correlation functions.
These three steps 
are iterated until a satisfactory statistical accuracy is obtained 
for the latter quantities.

The algorithm works with a finite number $M$ of walkers.   
Starting from the first walker, corresponding conventionally to the index 
$j=1$,  the basic steps of the algorithm are described below:
\begin{enumerate}
\item In the GF evolution, the exact propagator $e^{-H\Delta\tau}$ 
and the FN one
$e^{-H^{eff}\Delta\tau}$ are applied statistically for a given 
imaginary time interval
$\Delta\tau$. In practice this can be done by setting initially $\Delta\tau_{l}=
\Delta\tau$
and repeating the following steps until $\Delta\tau_{l}>0$:
\begin{enumerate}

\item Given the configuration of the walker, $x_{j}$,
the quantities $E_{x_j}$, ${\cal V}_{\rm sf} (x_j)$ and $H^{eff}_{x_{j},x_{j}}$ 
Eqs.~(\ref{elocal},\ref{signflip},\ref{heffdiag}) are
evaluated.  Then the interval $\Delta\tau_{d}$ during which the walker is 
expected to perform only diagonal moves (see Appendix \ref{app4}) 
is computed using the relation
$\Delta\tau_{d} = \min ( \Delta\tau_{l}, \ln{\xi}/ \pi_{d})$, where
$\xi$ is a random number between 0 and 1 and
$\pi_{d}= \lim_{\Lambda \to \infty} \Lambda \ln{p_{d}}=E_{x_j}-H^{eff}_{x_{j},x_{j}}$ 
according to Eq.~(\ref{pdiag}).

\item $\Delta\tau_{l}$ is updated 
$\Delta\tau_{l} \rightarrow \Delta\tau_{l}{-}\Delta\tau_{d}$ 
and the walker weights $(w_j,w^{eff}_j)$ are multiplied
respectively  by $e^{(-E_{x_j}-(1+\gamma)V_{sf}(x_j))\Delta\tau_{d}}$ and
$e^{-E_{x_j}\Delta\tau_{d}}$. Then
if $\Delta\tau_{l}>0$ a new configuration $x^{\prime}_{j}\neq x_{j}$ 
is chosen according to the probability table defined only by the normalized 
off-diagonal matrix elements of $p_{x^\prime,x_j}$, 
$${p_{x^\prime,x_j} \over \sum_{x^\prime \ne x_j}  p_{x^\prime,x_j}}~,$$ 
and  the weight $w_{j}$ is
multiplied by $s_{x^{\prime}_j,x_{j}}$ (\ref{sxprimex}).
The GF evolution then restarts from (a).
Otherwise, if $\Delta\tau_{l} = 0$ the GF evolution for the walker $j$
terminates and the algorithm proceeds for the next walker starting from (1). 
\end{enumerate}

\item
 After that  all the walkers $(w_j,w^{eff}_j,x_j)$ have been 
 propagated for the total imaginary time interval  $\Delta \tau$  the SR 
can be applied.  The  mixed averages  
$O^{k}_{x_j}=\langle \psi_{\rm G} | O | x_{j} \rangle/
\langle \psi_{\rm G} | x_{j} \rangle$ are computed 
at the end of such propagation  for the chosen set of operators $O^k$.
With these quantities  both 
$ \bar O_{eff}^k=\sum_j w_{j}^{eff} O^k_{x_j} /\sum w_{j}^{eff} $
and the covariance matrix $s_{k,k^\prime}$ in Eq.~(\ref{covmat})
are evaluated. By using the latter quantities in  the 
linear system (\ref{lineq}), 
the coefficients $\alpha_k$ are easily computed and
the table  $p_{x_j}$ is determined  according to Eq.~(\ref{defpxa}).
At this stage the reconfiguration procedure of the
walkers can be eventually performed,  i.e., the new $M$ configurations of the
walkers  are chosen among the old ones according to the probability
$|p_{x_j}|/\sum_{k} |p_{x_k}|$. 

\item The mixed averages of the physical observables $O^{k}_j$ and the quantity 
$$\frac{\sum_{k} w_{k}}{M} \,\frac{\sum_{k} |p_{x_{k}}|}{\sum_{k} p_{x_{k}}}~,$$
needed for the calculation of the statistical averages, are stored.
The walker weights are set to
$w_{j}={\rm sgn}\, p_{x_{j}}$ and $w_{j}^{eff}=1$,  and
the GF evolution can continue from step (1), 
starting again from the first walker.

\end{enumerate}

In the practical implementation of the algorithm the FN  dynamic
can be worked out at fixed $\gamma$, where $\gamma$ has to be a  non-zero number
otherwise  the exact GF could not be sampled 
(see Eqs.~(\ref{heffoff},\ref{heffdiag})). On the other hand
for $\gamma=0$ the FN is more accurate. A compromise is to work
with $\gamma=0.5$ fixed. Another choice is to implement few runs with
different non-zero $\gamma$ and try to extrapolate the results for $\gamma=0$,
 which should represent the most accurate calculation.
Typically this extra effort is not necessary because there is a very weak
dependence of the results upon $\gamma$. However the extrapolation to
$\gamma \to 0$ is an interesting possibility for the extension of the method
to continuous models, since, in this case,  there is no
practical way to cross the nodes with a variational FN approach (as shown in
Appendix \ref{app3} for the lattice case).

\section{ The limit of small $\Delta \tau$ and large number of walkers}

In this section some general properties of the  GFMCSR technique 
are discussed and explicitly  tested   on  the
$J_{1}{-}J_{2}$ Heisenberg model
\begin{equation}
\hat{H}=J{_1}\sum_{\langle i,j \rangle}
\hat{{\bf {S}}}_{i} \cdot \hat{{\bf {S}}}_{j}
+ J{_2}\sum_{\langle\langle i,j \rangle\rangle}
\hat{{\bf {S}}}_{i} \cdot \hat{{\bf {S}}}_{j}~~,
\label{j1j2ham}
\end{equation}
where ${\bf \hat{S}}_{i}$ are the  s-1/2 operators
sitting on the sites of a square lattice.
$J_{1}$ and $J_{2}$ are  the (positive) antiferromagnetic superexchange
couplings between nearest and next-nearest-neighbors pairs of spins
respectively.
In the following we will consider finite square clusters of $N$ sites with 
periodic boundary conditions.
We use the same guiding wavefunction of Ref.~\cite{letter} and report here 
some test results  useful to understand the crucial dependence of GFMCSR on  
the number of walkers $M$  and 
the frequency of the SR $\Delta \tau$
(the distance in imaginary time between two consecutive  SR).
In fact, after the selection of  a given number $p$ of  correlation functions 
in Eqs.~(\ref{srcon}), the 
results depend only on the number of walkers $M$ 
and the frequency of reconfiguration $\Delta \tau$. 
In the limit of large number of walkers, at fixed $p$,
the algorithm has the important 
property that the fluctuations of the coefficients $\alpha_k$ and 
$\bar O^k$ in Eq.~(\ref{defpx}) are  obviously vanishing, 
because they depend on ``averages'' 
of a very large number of samples of many different walkers, 
implying that these fluctuations  are  decreasing with  $1/\sqrt{M}.$ 
In this limit it is possible to recover an  important 
property of  the FN: {\em if the guiding wavefunction is exact, 
the FN averages $\bar O^k$ are also exact}. 
In fact suppose we begin to apply  the propagator $e^{-H \tau}$  
starting at $\tau=0$ from the 
exact sampling of the ground state 
$\psi_0$ determined  by FN with the exact guiding 
wavefunction $\psi_G=\psi_0$. 
Then at any Markov iteration $n$, before the SR is applied, 
both the mixed average correlation functions calculated with 
the FN weights 
$w^{eff}$ ($\langle O^k\rangle= { \sum_j w^{eff}_j O^k_{x_j} \over 
\sum_j w^{eff}_j } $) 
and the weights with arbitrary  signs 
$w$ ($\langle O^k\rangle= { \sum_j w_j O^k_{x_j} \over  
\sum_j w_j } $) sample statistically  the true quantum average
$\langle \psi_0 |O^k |\psi_0\rangle$.
If, for large $M$, we can neglect statistical fluctuations of these  averages,
 then by   
Eq.~(\ref{condition}) $\alpha_k=0$ and the SR algorithm 
just replace the weights $w_j$ (with sign problem) with the FN weights 
$w^{eff}_j$, which also sample  $\psi_0$ exactly if $\psi_G=\psi_0$. 
This means that 
the SR approach does not affect this important property of the FN, 
at least in the limit $M\to \infty$.

Another reason to work in the limit $M\to \infty$ is  the  following. 
In this limit it is not necessary to put in the SR conditions (\ref{condition})
operators $O^k$ that vanish for some symmetry that is satisfied both by the 
Hamiltonians $\bar H$ and the FN one $\bar H^{eff}$. 
In fact if the coefficients $p_{x_j}$ are defined in terms of operators $O^k$ 
that conserve the mentioned symmetries (e.g.  translation invariance, rotation 
by $90^0$  degree of the lattice etc.) by definition Eqs.~(\ref{srcon}) 
are satisfied  for all  the remaining  non-symmetric  operators 
which have vanishing expectation value  due to symmetry constraints
(such as e.g. an operator that changes sign for a rotation 
operation which is a symmetry of the Hamiltonians $\bar H$ and
 $\bar H^{eff}$). 
In this case  both sides of Eqs.~(\ref{srcon}) are  zero by such symmetry
constraints.
Moreover for $M\to \infty$ the statistical fluctuations are neglectable and 
for the same reason also Eqs.~(\ref{condition}) are  automatically satisfied 
with vanishing $\alpha_k$ for the above mentioned  non-symmetric operators.
In this limit it is therefore useless  to include non-symmetric operators 
in  the SR (\ref{condition}).

Finally it is interesting that in this important limit $M \to \infty$, 
within the assumption 
that we can neglect the fluctuations of $\alpha_k$ and  $\bar O^k_{eff}$, 
the SR depends only on the propagated states $\psi_n^{eff} (x) $ and 
$\psi_n (x)$.  
In fact given the state $\psi_n (x)$ and the FN one $\psi^{eff}_n (x)$, then 
the  state $\psi^\prime_n(x)$ after the SR will be 
\begin{equation} 
\begin{array}{rcl}
\psi^\prime_n (x)&=& C ( 1 +\sum_k \alpha_k ( O^k_x - \bar O^k_{eff} ) )  
\psi_n^{eff}  (x) \nonumber  \\
\psi^{eff ~ \prime}_{n}  (x) & = & |\psi^\prime_n (x)| 
\end{array}~,
\label{qm}
\end{equation} 
where now the $\alpha_k$ are uniquely determined by the conditions 
(\ref{srcon}), whereas the normalization constant 
$C={ \sum_x \psi_n(x) \over \sum_x  \psi_n^{eff} (x) }  $, and, finally,
$\psi^{eff \prime}_{n}$ replace the FN propagated state $\psi_n^{eff}$ 
after the SR (due to the condition $w^{eff \prime}_{j}=|w^\prime_j |$).
In this limit the dynamic described by the SR constraints is therefore perfectly defined 
and has a meaning, which can be computed even in an exact calculation 
without the Monte Carlo sampling. 

\begin{figure}
\centerline{\psfig{bbllx=50pt,bblly=250pt,bburx=510pt,bbury=600pt,%
figure=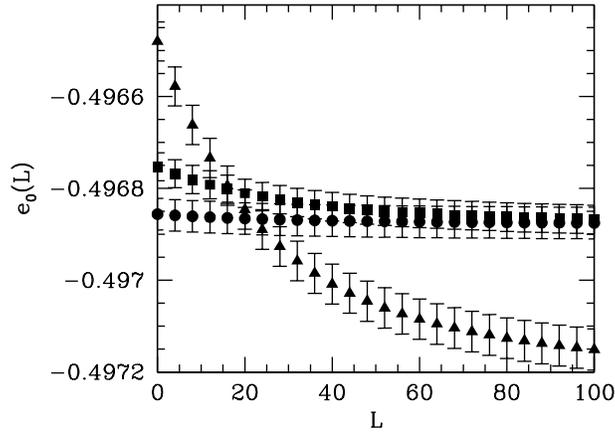,width=80mm,angle=0}}
\caption{\baselineskip .185in \label{corrfact}
Dependence on the number $L$ of correcting factors 
of the estimated ground state 
energy per site for $N=64$ and $J_{2}=0.5$ obtained with the 
GFMCSR technique ($\Delta \tau=0.01$)
with $M=$ 200 (triangles), 1500 (squares) and 10000 (circles). 
}
\end{figure}

\begin{figure}
\centerline{\psfig{bbllx=45pt,bblly=165pt,bburx=500pt,bbury=630pt,%
figure=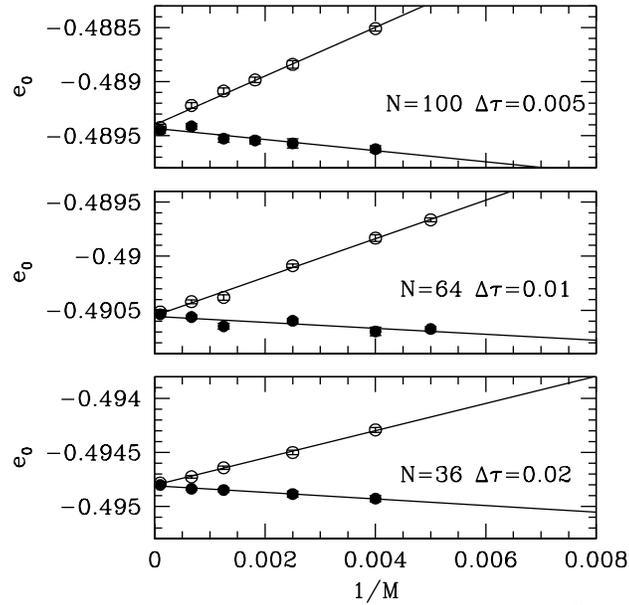,width=80mm,angle=0}}
\caption{\baselineskip .185in \label{varnw}
Ground state energy per site for $J_{2}=0.5$  obtained for different clusters
and different number of walkers. Empty dots are data obtained with zero
correcting factors while full dots refer to converged values in $L$.}
\end{figure}

The way the computed results depend on the number of walkers is shown 
in Fig.~\ref{corrfact}, as a function of the number of correcting factors.  
As it is evident for large number of walkers ($M\to \infty$) the correcting factors do not 
play any role and the estimate with minimum 
statistical error is obtained by simply ignoring the correcting factors. 
This is actually a common approach in GFMC, to consider a large number 
of walkers so that the bias of the finite walker population becomes 
neglectable, and typically decreasing as $1/M$ (see e.g. Fig.~\ref{varnw}). 
However from the picture it is also evident that for large enough $M$ 
the predicted results obtained by including or by neglecting 
the correcting factors are both consistent. 
The convergence to the $M\to \infty$ limit is however faster for the
first method. Thus the inclusion of the correcting factors $G^L_n$  in 
Eq.~(\ref{glfactors}), though increasing the error bars, may be useful to reach 
the $M\to \infty $ limit with a smaller number of walkers.
The fact that the two types of extrapolation to infinite $M$ -- the one  including
the correcting factors and the one neglecting them -- converge to the same 
value (see Fig.~\ref{varnw}) shows
that the theoretical limit when (\ref{qm}) holds 
can be reached with a reasonable number of walkers, much smaller than 
the dimension of the Hilbert space.  

The other parameter that affects the accuracy of the SR approach is 
the imaginary time distance  $\Delta \tau$ between two consecutive SR.
It is then natural to ask whether by increasing 
the frequency of the reconfigurations, one reaches a well defined 
dynamical limit for $\Delta \tau \to  0$. 
This is important since, due to the sign problem for large size the 
time interval $\Delta \tau$ {\em has } to be decreased at least by a factor
inversely proportional to the system size, because 
the average walker sign vanishes exponentially  $\sim e^{- \Delta_s \tau } $ 
with an exponent $\Delta_s$  which diverges with the system size. 
Different calculations, performed for different sizes can be compared 
only when the finite $\Delta \tau $ error (the difference between $\Delta \tau\to 0$ 
and finite $\Delta \tau$) is neglectable.

\begin{figure}
\centerline{\psfig{bbllx=55pt,bblly=255pt,bburx=500pt,bbury=590pt,%
figure=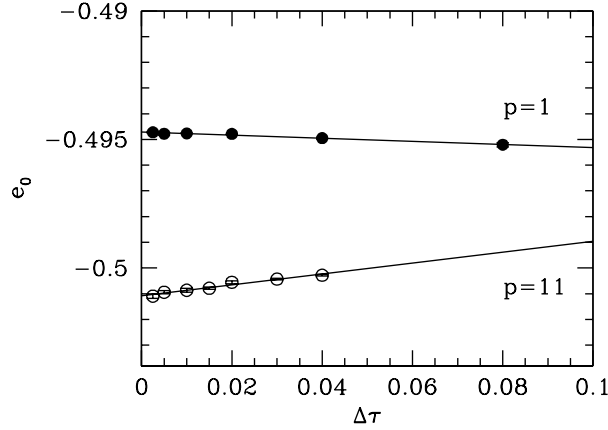,width=80mm,angle=0}}
\caption{\baselineskip .185in \label{dtlimit}
Dependence of the ground state energy per site on the imaginary time step
$\Delta \tau$ obtained for $J_{2}{=}0.5$ and $N=36$ with the GFMCSR technique
by using in the SR the energy (full dots), all $S^z(q)$, the spin square
and the order parameter $m^{\dagger 2}$ (empty dots). The number of walkers was
fixed to $M=10000$, so that the finite-$M$ bias can be neglected on this scale.
The lower horizontal axis coincides with the exact diagonalization result.}
\end{figure}

As shown in Fig.~\ref{dtlimit}, whenever the simulation 
is stable for $\Delta \tau\to 0$ 
the limit $\Delta \tau \to 0$ 
can be reached with a linear extrapolation. This 
property can be easily understood 
since in the limit of large number of walkers  the variation of the average 
correlation functions Eq.(\ref{glfactors}) both for the FN dynamic and the 
exact dynamic in a time interval between two consecutive SR  
differ clearly by $O(\Delta \tau) $. 
 
\begin{figure}
\centerline{\psfig{bbllx=40pt,bblly=260pt,bburx=510pt,bbury=590pt,%
figure=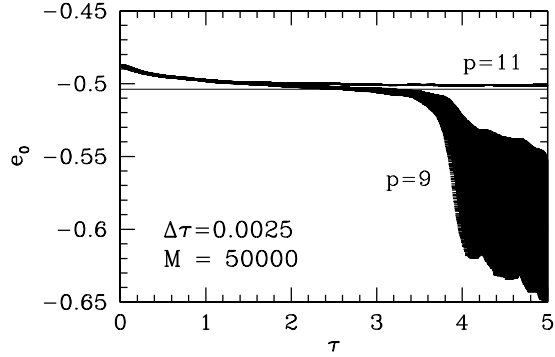,width=80mm,angle=0}}
\caption{\baselineskip .185in \label{instab}
Stable (upper curve) and unstable (lower curve) imaginary time evolution of 
the GFMCSR estimates of the ground energy per site 
for $J_{2}=0.5$ and the $N=36$ cluster. The horizontal line indicates the exact
result.}
\end{figure}

In order to show more clearly how the method is working and systematically 
correcting the FN we have implemented  a slightly different 
but more straightforward ``Release Nodes technique''\cite{release}. 
We first apply the standard FN (with $\gamma=0$, see Eq.~(\ref{heffoff}))
for a given number of walkers $M$ and for long simulation time. We   
store the  $M$-walkers configurations, after some equilibration 
at time interval large enough to allow uncorrelated and independent 
samples of the FN ground state. 
In the second step we recover each of these $M$-walker configurations
and apply GFMCSR  for a fixed imaginary time $\tau$, so that we can see 
how the energy expectation value evolves from the FN to a more 
accurate determination.  
Typically one obtains a reasonable behavior for these curves that always 
coincides with the exact dynamic in the initial part where an exact sampling 
of the sign is possible. 
However for large imaginary time, and exceedingly small $\Delta \tau$ and 
large number of walkers, some instability may occur 
leading to results clearly off, as shown in Fig.~\ref{instab}. 
In this case the reason of the instability is 
due to the fact that the correlation functions 
$S^z(q)={ 1\over N^2} \sum_{i,j} S^z_i S^z_{j} e^{i q (i-j) } $ 
which we have used in the SR ($p=9$)\cite{letter}, introduce some 
uncontrolled fluctuations for the momentum  $Q=(\pi,\pi)$ relevant for 
the antiferromagnetic order parameter. If we include in the SR 
technique also the spin  isotropic operator corresponding 
to the order parameter 
$m^{\dagger 2}= { 1\over N^2} \sum_{i,j} \vec S_i  
\cdot \vec S_{j} e^{i Q (i-j) } $ and the total spin square ($p=11$)
this instability disappears  (see Fig.\ref{instab}, 
stable results, not shown in the picture, are obtained
even without the total spin square, i.e. with  $p=10$).
This is a reasonable effect since the order parameter 
has important fluctuations in all spin directions.

\section{Conclusions}

In this paper we have tried to describe in detail  
a  recently proposed  technique GFMCSR, 
that allows to work within a controlled accuracy with the ground state energy 
and with related mixed average correlation functions even  for models where  
the conventional  Quantum Monte Carlo technique cannot be used  for   
the well known sign problem.

This method is rather general,  in principle 
convergence is achieved within an   arbitrary  
accuracy if a sufficiently large number $p$ of correlation functions is 
constraint to be equal before and after the SR, 
the basic statistical step used to  stabilize the sign problem
instability.
In order to minimize the number $p$ of correlation functions used in the SR,
one is limited to use 
an empirical approach, based on physical intuition, and/or by comparison 
with exact results obtained at finite size with the exact diagonalization
technique. Typically the fundamental ingredient that we have found important 
for strongly correlated Hamiltonians is the  ``locality''. The most  useful  
correlation functions are the short range ones  
contained in the Hamiltonian $H$.
A more successful  example is the application of the method 
to the  Heisenberg model on the triangular lattice\cite{luca} 
where a remarkable accuracy is obtained by including also the short 
range correlation functions generated by the application 
of the square Hamiltonian. Here we report a table (see Table I) 
with all the values of the ground state energy per site, 
the total spin square and
the antiferromagnetic order parameter $m^{\dagger 2}$ 
obtained with VMC, FN and GFMCSR 
(for two different $p$), up to $N=108$. 
This method to increase systematically $p$, 
by including in the SR the short range correlation functions 
generated by $H, H^2 \cdots$, 
does not seem general enough. For instance it does not 
work for the $J_1-J_2$ Heisenberg model where the inclusion of long range 
operators in the SR Eqs.~(\ref{srcon}) such as the spin-spin 
correlation function  $S^z_i S^z_j$ at large distance $|i-j|$ is  
crucial to improve the accuracy of the method, whereas the short range ones
generated by $H^2$ do not give any significant improvement. 

\begin{figure}
\centerline{\psfig{bbllx=65pt,bblly=300pt,bburx=510pt,bbury=590pt,%
figure=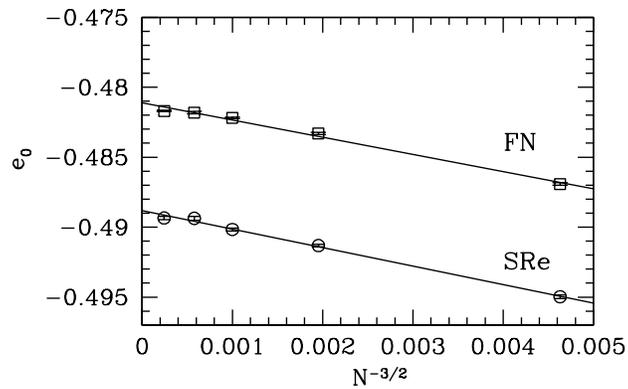,width=80mm,angle=0}}
\caption{\baselineskip .185in \label{sizecon}
Finite size scaling of the GS energy per site  for $J_{2}=0.5$ 
obtained with the FN and GFMCSR technique applied reconfiguring
the Hamiltonian (SRe).
}
\end{figure}

Similarly to FN the GFMCSR is size-consistent   
(see Fig.~\ref{sizecon}). At fixed  $p$  
a given accuracy is expected in the average correlation functions, 
accuracy  which  looks 
weakly dependent on the system size and  
different from the variational guess even in the thermodynamic limit. 
This is a very important property of the present algorithm 
because the stability of the average sign at fixed $p$ 
allows a  {\em polynomial}  complexity of the algorithm as a function of 
the system size. The algorithm, however, is typically a large factor
($\simeq 100$) more expensive than the standard FN as far as the 
computational time is concerned, for a given statistical error on correlation
functions. 

Until now the method has been extended rather successfully to several models:
the mentioned $J_1{-}J_2$ and triangular lattice  Heisenberg models, 
the $t-J$ model\cite{becca} and preliminary results show that 
similar improvement of the standard FN can be obtained also for the 
Hubbard model \cite{capone}.  
In the latter case it is worth to mention that a different approach, 
the Constrained Path Monte Carlo\cite{gubernatis} (CPMC)
represents also a very good remedy 
for the sign problem disease at least for intermediate coupling ($U/t \le 8$).
On the other hand different schemes to get rid of the ''sign problem'' for 
continuous systems were previously proposed and successfully applied to 
small electron systems.\cite{kalos91}

Although the GFMCSR  is far from being the definite solution of the 
sign problem in the Monte Carlo simulation, it certainly represents  
an interesting  possibility to alleviate this instability even for 
large system sizes.  
Its extension to continuous systems and also to 
CPMC is indeed straightforward, even though,  
in these cases,  the possibility to cross the nodal surface in a variational 
way (see Appendix \ref{app2})  is not possible at present. 

\acknowledgements
We acknowledge useful discussions with  M. Calandra, F. Becca, G. Bachelet 
and J. Carlson. 
This work was partially supported by INFM (PRA HTSC and LOTUS) and 
by ``MURST-Progetti di ricerca di rilevante interesse nazionale''.

\appendix
\section{Property of a stochastic matrix }
\label{app1}
In this appendix we remind some properties of a stochastic matrix
 $p_{x^\prime,x}$. 
The stochastic matrices are square matrices that  have  all non-negative 
matrix elements  $p_{x^\prime,x}$ and  satisfy the normalization condition
\begin{equation} \label{np}
\sum_{x^\prime} p_{x^\prime,x} = 1~, 
\end{equation}
for each column  matrix index $x$.
We assume also that the number of row and column indices are finite 
and that each index  $x$ is connected to any other $x^\prime$  
by at least one  sequence $p_{x^\prime,x_1} p_{x_1,x_2} \cdots 
p_{x_N,x} $ of non-zero matrix elements of $p$.

The stochastic matrices are generally  non-symmetric and   their 
eigenvalues may be also complex. For each eigenvalue  there exist a left 
$ \sum_{x^\prime} \psi_L (x^\prime)  p_{x^\prime,x}  = \lambda \psi_L(x)$  
and a corresponding  right eigenvector $\sum_x p_{x^\prime,x} 
\psi_R(x) =\lambda \psi_R(x^\prime)$.
A very simple left eigenvector is the constant one $\psi_L(x)=1$, 
that by property (\ref{np}) has eigenvalue $\lambda=1$.
We will show in the following that this is actually the maximum eigenvalue 
because:
\noindent  {\em i}) {\em to each right eigenvector $\psi_R(x)$ of $p$ corresponds 
an eigenvalue $\lambda$, which is bounded by one $|\lambda|<1$ }.

In fact,  be $\psi_R(x)$ a generic (complex or real) right eigenvector of $p$
$$  \lambda \psi_R(x^\prime) =  \sum_x p_{x^\prime,x} \psi_R(x)~,  $$
by taking the complex modulus of both sides 
of the previous  equation and summing over $x^\prime$ we obtain
$$ |\lambda | \sum_{x^\prime} |\psi_R(x^\prime)| = 
\sum_{x^\prime} |  \sum_x p_{x^\prime,x} \psi_R(x) | \le 
\sum_{x}   \sum_{x^\prime}  p_{x^\prime,x} |\psi_R(x) |= \sum_x |\psi_R(x) |~,$$
where in the above  inequality we have interchanged the summation indices 
and used the elementary bound for the complex modulus 
$|\sum_x  z_x | \le \sum_x  |z_x| $ for  arbitrary numbers $z_x= p_{x^\prime,x} \psi_R(x)$.
This immediately gives:
$$|\lambda| \le 1~.$$
Obviously the equality sign holds if, for each $x$,
$|\sum_x  z_x | = \sum_x  |z_x| $, which implies  that 
given a right eigenvector 
with maximum eigenvalue $\lambda=1$, the real positive definite 
vector $|\psi_R(x)|$ is also a right eigenvector with maximum eigenvalue.

 Now we will show that: 
\noindent  {\em ii}) {\em the maximum right eigenvector is unique}. 
In fact suppose that there are two right eigenvectors $\psi_1$ and $\psi_2$ 
with $\lambda=1$, then  by linearity also $\psi_1-\alpha \psi_2$ is 
a right eigenvector with $\lambda=1$ and the complex 
 constant $\alpha$ can be chosen  to give $\psi_1-\alpha \psi_2=0$ for 
a given index $x_0$.
On the other hand using the property derived previously also 
$| \psi_1(x)  -\alpha \psi_2 (x)| $  is a right maximum eigenvector
 that vanishes  for $x=x_0$.
Using iteratively the definition of a right eigenvector 
$$\sum_{x} p_{x^\prime,x} |\psi_1(x)-\alpha \psi_2(x)| = | \psi_1(x^\prime)
-\alpha \psi_2(x^\prime) |~, $$
starting from $x^\prime=x_0$, we arrive easily to derive that for all the 
index $x$ connected to $x_0$ by non-zero sequence of 
matrix elements   $p_{x_0,x_1} p_{x_1,x_2} \cdots
p_{x_N,x} $  
$$ |\psi_1(x)-\alpha \psi_2(x)| =0.$$
Since by hypothesis all the possible indices are connected to $x_0$ 
by at least one such a sequence, 
we derive $\psi_1 = \alpha \psi_2$, which means that $\psi_1$ and 
$\psi_2$ are the same eigenvector, which contradicts the initial 
hypothesis. Thus the maximum right eigenvector is unique. 

Collecting the above properties, the maximum right eigenvector $\psi_R(x)$ 
of a stochastic matrix can be chosen real and positive.  
Then it is simple to show that the iteration of the stochastic matrix
$$ p^n \psi_T$$
converges for large $n$ to this maximum right eigenvector 
with an exponentially decreasing error $\propto \gamma^n$, with $\gamma < 1$ 
being the modulus of largest eigenvalue of $p$, different from the maximum one.

\section{ Proof of the Upper bound}
\label{app2} 
Here we follow the paper \cite{ceperley1} to prove rigorously the 
upper bound property of the ground state energy  for $H^{eff}$. 
We want to show that the prescription given  in 
Eqs.~(\ref{heffoff},\ref{heffdiag}) 
for $ H^{eff}$ leads to an upper bound for the ground state
energy of $ H$. 
When importance sampling is used it is important to change 
slightly the definition of the sign-flip term as in (\ref{signflip}):
\begin{equation}\label{signguiding}
 {\cal V}_{\rm sf} (x) =
\sum\limits_{\psi_G (x^\prime) H_{x^\prime, x}/\psi_G(x) 
 > 0~  {\rm and}~ x^\prime \ne x  }
\psi_G (x^\prime) H_{x^\prime,x}/\psi_G(x)~. 
\end{equation}
 
We now take {\em any\/} state
\begin{equation} |\psi\rangle = \sum_x |x\rangle\psi(x),
\end{equation}
and we compare its energy with respect to $ H$ and to $
H^{eff}$:
\begin{equation} \nonumber
\Delta E = \langle\psi|( H^{eff}- H)|\psi\rangle~.
\end{equation}

$\Delta E$ can be written explicitly in terms of the matrix elements of
$H$, using the definitions given in
Eqs.~(\ref{heffoff},\ref{heffdiag},\ref{signguiding})
\begin{equation}\label{doublesum}
\Delta E = (1+\gamma)   \sum_x\psi(x)^\ast\left[
\sum_{x^\prime}^{\text{sf}}
H_{x,x^\prime} \frac{\psi_G (x^\prime)}
{\psi_G( x)}\psi(x)
- \sum_{x^\prime}^{\text{sf}} H_{x,x^\prime} \psi(x^\prime)
\right],
\end{equation}
where the notation $sf$ indicates conventionally the summation over
the off-diagonal elements such that 
$\psi_G(x) H_{x,x^\prime}/\psi_G(x^\prime) > 0 $.
In this double summation each pair of configurations $x$ and $x^\prime$ occurs
twice. We combine these terms and rewrite (\ref{doublesum}) 
as a summation over pairs:
\begin{equation}
\Delta E = (1+\gamma)   \sum_{(x,x^\prime)}^{\text{sf}}
H_{x,x^\prime} \left[
\left|\psi(x)\right|^2\frac{\psi_G (x^\prime)}{\psi_{G}(x)} +
\left|\psi(x^\prime)\right|^2\frac{\psi_G (x)}
{\psi_G (x^\prime)} - \psi(x)^\ast\psi(x^\prime) -
\psi(x^\prime)^\ast\psi(x) \right] .
\end{equation}
Denoting by $sH(x,x^\prime)$ the sign of the matrix element $
H_{x,x^\prime}$, and using the fact that for all terms in this
summation the condition 
$\psi_G (x^\prime)  H_{x^\prime,x} \psi_G (x) > 0 $ 
is satisfied, we can finally write $\Delta E$ as
\begin{equation} \label{deltaEfinal}
\Delta E =  (1+\gamma)  \sum_{(x,x^\prime)}^{\text{sf}}\left|
H_{x,x^\prime} \right|
\left|\psi(x)\sqrt{\left|\frac{\psi_{G}(x^\prime)}
{\psi_{G}(x)}\right|} -
sH(x,x^\prime)\psi(x^\prime)\sqrt{\left|\frac{\psi_{G}(x)}
{\psi_G (x^\prime)}\right|}\right|^2 .\end{equation}
Obviously, $\Delta E$ is positive for any wavefunction $\psi$. Thus the
ground state energy of $H^{eff}$ is an upper bound for the
ground state energy of the original Hamiltonian $ H$.

Now the GFMC method can calculate the exact ground state energy
$E^{eff}_0$ and wavefunction $\psi^{eff}$ of $H^{eff}$, without 
any sign problem.  Hence:
$E^{eff}_0 \ge \langle\psi^{eff} | H | \psi^{eff}\rangle \ge E_0$,
where the second inequality follows from the usual variational principle.
We conclude therefore that the FN energy is an upper bound 
to the true ground state energy.
One can easily verify that $\langle \psi_G |  H|\psi_{G}\rangle =
\langle \psi_G | H^{eff}|\psi_{G}\rangle$, and thus one can be sure that
the GFMC procedure improves on the energy of the guiding  wavefunction:
$E^{eff}_0 \le \langle\psi_{G} | H^{eff}
|\psi_{G}\rangle = \langle\psi_{G} | H | \psi_{G}\rangle$.

Note that the standard ``lattice FN'' approach \cite{ceperley1} is 
obtained for the particular parameter $\gamma=0$. 

\section{Proof of existence and unicity of solution for the reconfiguration}
\label{app3}
 In this appendix we prove that given the $p+1$ SR conditions (\ref{condition})
the elements of the table $p_{x_j}$ are uniquely determined for each 
walker configuration $(\underline{w},\underline{x})$.

We define here the quantity
\begin{equation}
v^k_j = (O^k_{x_{j}}- \bar O^{k}_{f})~, 
\end{equation}
for each configuration $j$, where 
$ \bar O^k_f= { \sum_j w^{f}_j O^k_{x_j} \over
\sum_j w^{f}_j }$ is the average value over the reference weights, $w_{j}^f$,
of the operator considered, labeled by the number $k$.  
The reference weights $w^{f}_j$ are restricted to be strictly positive
but arbitrary functions of all the FN weights $\{w^{eff}_j\}$ 
the exact weights $\{w_j\}$ and the configurations $\{x_j\}$.
It is easy to show that, in order  that 
\begin{equation} \label{defpxa}
p_{x_j} = w^f_j (  1+ \sum_k \alpha_k v^k_j) 
\end{equation}
allows to satisfy the SR conditions (\ref{srcon}), 
it is sufficient that $\alpha_k$ are determined by 
the simple linear equation
\begin{equation}\label{lineq} 
\sum_{k^\prime} s_{k,k^\prime} \alpha_{k^\prime} = { \sum_j w_j v^k_j \over
\sum_j  w_j }~, 
\end{equation}
where 
\begin{equation}\label{covmat}
s_{k,k^\prime} = { \sum_j w^f_j v^k_j v^{k^\prime}_j \over \sum_j w^f_j } 
\end{equation}
is the covariance matrix of the operators $O^k$ over the reference weights 
$w^f_j$.
The solution to (\ref{lineq}) is possible if the determinant of $s_{k,k^\prime}$
is non-vanishing.
Since $s$ represents an overlap matrix defined with a non-singular scalar
product  $  \langle v^k | v^{k^\prime} \rangle = { \sum_j w^f_j v^k_j v^{k^\prime}_j \over
\sum_j w^f_j }$ as $w^f_j$ are positive, its determinant is always non-zero 
provided the vectors $v^k$ are linearly independent. 
Thus, in the latter case, the solution to (\ref{lineq}) exists and is unique.

On the other hand suppose that among the $p$ vectors $v^k$ only $p^\prime < p$ 
are linearly independent. Thus the remaining $p-p^\prime$ vectors can be
written as linear combination of  $p^\prime$ linearly independent ones 
(henceforth we assume that these linearly independent vectors are labeled 
by the consecutive indices $k=1,\cdots, p^\prime$)
\begin{equation} \label{vrest}
v_j^{k^\prime} = \sum\limits_{k=1}^{p^\prime} x^{k^\prime}_k v^k_j~,
\end{equation}
for $k^\prime>p^\prime$, where $x^{k^\prime}_k$ are suitable coefficients.
The same previous considerations allow to satisfy the first $p^\prime$ SR
conditions  as for Eq.~(\ref{lineq})   a unique solution   exists if we restrict
all  the sums for $k, k \le p^\prime$, and  $p_{x_j}$ is  determined
only by the first $p^\prime$ linearly independent vectors in (\ref{defpxa}).
With the determined  $p_{x_j}$ it is obvious that    
\begin{equation} \label{refcon}
{ \sum_j p_{x_j} v^{k}_j \over \sum_j p_{x_j} } = { \sum_j w_j
v^{k}_j \over \sum_j w_j } 
\end{equation}
is verified for $k=1,\cdots, p^\prime$. 

 On the other hand  we can easily show that all the remaining
SR conditions (\ref{refcon})  for $k^\prime > p^\prime$ are identically
satisfied. In fact, in this case the LHS of Eq.~(\ref{refcon}) can be
manipulated as follows, using definition (\ref{vrest})
\begin{equation}
{ \sum_j p_{x_j} v^{k^\prime}_j \over \sum_j p_{x_j} }  = 
\sum_k x^{k^\prime}_k \left( { \sum_j v^k_j p_{x_j} \over \sum_j  p_{x_j}
}\right)= \sum_k x^{k^\prime}_k \left( { \sum_j v^k_j w_j \over \sum_j w_j
}\right)={ \sum_j v^{k^\prime}_j w_j \over \sum_j w_j }~,      
\end{equation}
where in the intermediate steps we have used that 
$$\left( { \sum_j v^k_j p_{x_j} \over \sum_j p_{x_j} }\right)=
\left( { \sum_j v^k_j w_j \over \sum_j w_j }\right)~,$$ 
for the first $k \le p^\prime$ conditions.
Thus the SR conditions determine uniquely $p_{x_j}$ in any case and this 
conclude the important statement of this Appendix.
\subsection{Remark}
\label{subapp3}
With the above definitions it is also possible to show that $p_{x_j}$ 
remains unchanged for any linear transformation of the operator set.
Namely, suppose we consider the new operators
\begin{equation} \label{transf}
\tilde O^{k^\prime} = \sum_k L_{k^\prime,k} O^k + \beta_{k^\prime}
\end{equation}
in the SR conditions, where the real matrix $L$ is assumed to have 
non-vanishing  determinant.
Within this assumption it is simple to show that  $p_{x_j}$ will remain
unchanged. 

In fact the new set of operators will define a new covariance matrix 
between the new vectors 
\begin{equation} \label{lv}
\tilde v^{k^\prime}_j = \sum_{k} L_{k^\prime,k} v^k_j~, 
\end{equation}
i.e., $\tilde v = L v$, 
$\tilde s = L s L^T$, where $L^T $ is the transposed of $L$ and the set of
new equations 
$$ \sum_{k^\prime} \tilde s_{k,k^\prime}  \tilde \alpha_{k^\prime} = { \sum_j w_j 
\tilde v^k_j \over \sum_{j} w_j}$$ 
is obviously satisfied by 
\begin{equation} \label{sola}
\tilde \alpha = (L^{-1})^T \alpha~, 
\end{equation}
 where $\alpha$ 
is the solution of the SR conditions before the transformation (\ref{transf}).
Whenever the number $p^\prime$ of  linearly independent $v^k$ is less than $p$, 
also  the number of linearly independent $ \tilde v^k$ will be $p^\prime$ as 
$L$ is non-singular. 
The solutions $\alpha $ and $\tilde \alpha$, as described
previously, refer therefore to the first $p^\prime$ components, and all the
matrix involved, such as $\tilde L $ and $\tilde s$  are in this case  
restricted to this subspace.  
 
Then, by Eq.~(\ref{sola}) and Eq.~(\ref{lv}),  it easily follows that the new
coefficients $\tilde p_{x_j} = w^f_j ( 1 + \sum_k \tilde \alpha_k  \tilde v^k_j
) =w^f_j ( 1 + \sum_k  \alpha_k   v^k_j
) = p_{x_j}$, which finally proves the statement of this remark.  

\section{The limit $\Lambda \to \infty$ for the power method}
\label{app4}

The constant $\Lambda$, which defines the 
the Green function $G_{x^\prime,x}=\Lambda
\delta_{x^\prime,x} -H_{x^\prime,x}$ and the FN one $G^{eff}$ (\ref{geff}) 
has to be taken large enough to  determine that all the diagonal 
elements of $G^{eff}$ are  non-negative (by definition the off-diagonal ones of
$G^{eff}$  are always non-negative). This requirement often determines a very large 
constant shift which increases with  larger size and is not known a priori.
The trouble in the simulation may be quite tedious, as if for the chosen
$\Lambda$ a negative diagonal element is found for $G^{eff}$, one needs to
increase  $\Lambda$ and  start again with a completely new simulation.
The way out is to work with exceedingly  large $\Lambda$, but this may slow
down the efficiency of the algorithm as in the stochastic matrix
$p_{x^\prime,x}$ the probability to remain in the same configuration $p_d$ 
may become very close to one
\begin{equation}\label{pdiag}
p_d= { \Lambda - H_{x,x} -(1+\gamma) {\cal V}_{\rm sf} (x)  \over \Lambda - E_x } 
\end{equation}
where ${\cal V}_{\rm sf}$ is given in Eq.~(\ref{signflip}) and $E_x$ is 
the local energy  Eq.~(\ref{elocal})  that do not depend on $\Lambda$ given the 
configuration $x$.

Following Ref.~\cite{ceperley} the problem of working with large $\Lambda$ 
can be easily solved with no loss of efficiency. 
We report this simple idea applied  to our particular algorithm at
fixed number of walkers. 
If $\Lambda$ is large it is possible to take a large value of $k_p$ 
(of order $\Lambda$)  iterations between two consecutive reconfigurations,
because in most iterations  the  configuration $x$ is not changed.
The idea is that one can determine a priori, given $p_d$ what is the 
probability $t(k)$ to make $k$ diagonal moves  before the first acceptance of a
new configuration  with $x^\prime \ne x$.
This is given by $t(k)=p_d^k (1-p_d)$ for $k=0,\cdots, n_l-1$ and
$t(n_l)=p_d^{n_l}$ if no off-diagonal moves are accepted during the $n_l$
trials that are left to complete the loop without reconfigurations.

It is a simple exercise to show that, in order to sample $t(k)$  one needs  one 
random number $0< \xi < 1 $, so that  
 the stochastic integer number $k$ can be  computed by the simple formula 
\begin{equation}\label{evalk}
k = \min ( n_l,[ { \ln \xi \over \ln p_d } ] )~,
\end{equation}
where the brackets indicate the integer part. 
During the $k_p$ iterations one can iteratively apply this formula by
bookkeeping the number of iterations $n_l$ that are left to complete the loop
without reconfigurations.
At the first iteration  $n_l=k_p$, then $k$ is extracted using (\ref{evalk}), 
and the weights $(w,w^{eff})$ of the walker are updated according to $k$ diagonal moves
and  if $k< n_l$ 
a new configuration is extracted randomly according  to the off-diagonal matrix
elements of $p_{x^\prime,x}$. The weights are correspondingly updated 
for this off-diagonal move, and finally, if $k<n_l$, $n_l$ is changed to  $
 n_l -k -1$, so that one can  continue to use Eq.~(\ref{evalk})  until all the
$k_p$ steps are executed  for each walker.

The interesting    thing of this method is that it can be readily generalized
for  $\Lambda \to \infty$ by increasing $k_p$ with $\Lambda$, namely 
 $k_p = [\Lambda \Delta \tau]$, where $\Delta \tau$ represents now exactly the imaginary time 
difference between two consecutive reconfigurations where the exact propagator 
$e^{ - H \Delta \tau} $ or $e^{ - H^{eff} \Delta \tau} $ is applied statistically. 

\begin{table} 
\label{tab}
\begin{tabular}{dcccccc}
               &      N        &   VMC       &   FN        &  SR($p=2$)  &  SR($p=7$) & Exact \\
\tableline
$e_{0}$        &     12        & -0.5981     & -0.6083(1)  & -0.6085(1)  & -0.6105(1) &-0.6103\\ 
               &     36        & -0.5396     & -0.5469(1)  & -0.5534(1)  & -0.5581(1) &-0.5604 \\
               &     48        & -0.5366(1)  & -0.5426(1)  & -0.5495(1)  & -0.5541(1) & \\
               &    108        & -0.5333(1)  & -0.5387(1)  & -0.5453(1)  & -0.5482(1) & \\
\tableline
$S^2_{tot}$    &     12        &  0.235      & 0.0111(2)  &  0.006(4)   & -0.002(4)  & 0.00\\ 
               &     36        &  1.71       & 1.20(1)    &  0.65(1)    &  0.02(1)   & 0.00\\
               &     48        &  2.55(1)    & 2.12(2)    &  1.44(1)    &  0.23(3)   & 0.00\\
               &    108        &  6.36(4)    & 5.66(3)    &  4.35(4)    &  2.7(1)    & 0.00\\
\tableline
$m^{\dagger 2}$&     12        &  0.9241     & 0.9286(1)  &  0.9210(2)  & 0.9132(6)  & 0.9109\\ 
               &     36        &  0.7791     & 0.7701(4)  &  0.7659(2)  & 0.7512(3)  & 0.7394\\
               &     48        &  0.7496(3)  & 0.7243(5)  &  0.7177(2)  & 0.7080(5)  & \\
               &    108        &  0.6338(7)  & 0.6182(4)  &  0.6040(3)  & 0.5836(5)  & \\
\end{tabular}
\caption{Variational estimate (VMC) and mixed averages (FN, SR and Exact)
of the ground energy per site, the total spin square and the order parameter 
for the triangular Heisenberg antiferromagnet for various system sizes.
SR data are obtained using the short range correlation functions generated
by $H$ ($p=2$) and $H^2$ ($p=7$) reported in Ref.~\protect\cite{luca}.
All the  values reported in this table are  obtained with large enough 
$M$ and $1/\Delta \tau$,  practically converged in the limit
 of $\Delta \tau \to 0$ and $M\to infty$.
}
\label{mixedav}
\end{table}

\end{document}